\documentclass[aps,prb,twocolumn,groupedaddress,floatfix]{revtex4}

\usepackage{graphicx}
\usepackage{subfigure}

\def\be{\begin{equation}}
\def\ee{\end{equation}}
\def\bea{\begin{eqnarray}}
\def\eea{\end{eqnarray}}

\begin{document}


\title{Coulomb screening and electronic instabilities of 
small-diameter (5,0) nanotubes}


\author{J. Gonz{\'a}lez$^{1}$ and E. Perfetto$^{1,2}$}
\affiliation{$^{1}$Instituto de Estructura de la Materia.
        Consejo Superior de Investigaciones Cient{\'\i}ficas.
        Serrano 123, 28006 Madrid. Spain.\\
$^{2}$Istituto Nazionale di Fisica Nucleare - Laboratori
Nazionali di Frascati, Via E. Fermi 40, 00044 Frascati, Italy.}


\date{\today}

\begin{abstract}
We investigate the instabilities that may lead to the breakdown
of the Luttinger liquid in the small-diameter (5,0) nanotubes,
paying attention to the competition between the effective interaction 
mediated by phonon-exchange and the Coulomb interaction.
We use bosonization methods to achieve an exact treatment of the
Coulomb interaction at small momentum-transfer, and apply next
renormalization group methods to analyze the low-energy behavior of 
the electron system. This allows us to discern the growth of several 
response functions for charge-density-wave modulations and for 
superconducting instabilities with 
different order parameters. We show that, in the case of single
nanotubes exposed to screening by external gates, the Luttinger
liquid is unstable against the onset of a strong-coupling phase with
very large charge-density-wave correlations. 
The temperature of crossover to the new phase depends crucially on
the dielectric constant $\kappa $ of the environment, ranging from
$T_{c}\sim 10^{-4} \; {\rm K}$ (at $\kappa \approx 1$) up to a
value $T_{c} \sim 10^{2} \; {\rm K}$ (reached from $\kappa \approx
10$). The physical picture is however different when we consider
the case of a large array of nanotubes, in which there is a 
three-dimensional screening of the Coulomb interaction over distances much larger than 
the intertube separation. The electronic instability is then 
triggered by the divergence of one of the charge stiffnesses in the 
Luttinger liquid, implying a divergent compressibility and the 
appearance of a regime of phase separation into spatial regions 
with excess and defect of electron density.

\end{abstract}

\maketitle

\vspace*{1cm}
\section{Introduction}

During the last decade, carbon nanotubes have shown a great 
potential for the development of molecular electronics. 
This comes from their versatile electronic 
properties, which depend on variables like the helicity of the 
tubule, the temperature, or the contacts used in
the experiments. It is well-known for instance that the carbon
nanotubes may have semiconducting or metallic behavior, depending
on the roll-up direction of the wrapped hexagonal carbon 
lattice\cite{saito}. This has opened the way for the construction
of intramolecular junctions behaving as electronic devices at the
nanometer scale\cite{yao}.

Given their reduced dimensionality, the metallic nanotubes 
behave as strongly correlated electron systems, with different 
regimes depending on the energy scale. At room temperature, for
instance, the strong Coulomb repulsion prevails in nanotubes of
typical radius, driving the electron system to a state with the
properties of the so-called Luttinger liquid, characterized by 
the absence of electron quasiparticles at the Fermi level. This 
feature manifests in the power-law dependence of observables such 
as the tunnelling density of states, whose suppression at low 
energies has been actually observed in measurements of the 
conductance in the carbon nanotubes\cite{yao,exp}.

At low temperatures, the behavior of the carbon nanotubes 
depends on the contacts used in the transport 
measurements. When the contacts are not highly transparent, 
there is in general a suppression of the zero-bias conductance 
and the differential conductivitiy, characteristic of the 
Coulomb blockade regime. On the other hand, there have 
been experiments using high-quality contacts where it has 
been possible to observe superconducting correlations in the 
carbon nanotubes\cite{kas,marcus}. The most remarkable signature 
has been the appearance of supercurrents in nanotube ropes 
suspended between superconducting electrodes\cite{kas}. This 
feature may be understood as a consequence of the proximity 
effect, by which Cooper pairs are formed in the nanotubes near 
the superconducting electrodes\cite{th1}. Moreover, 
superconducting transitions have been measured in ropes 
suspended between metallic, nonsuperconducting 
electrodes\cite{sup,priv}. These observations imply the 
existence of a regime with a relevant attractive component in 
the electron-electron interaction, coming presumably from the 
exchange of phonons between electronic currents\cite{th2}.

>From a theoretical point of view, it is well-known that 
the Luttinger liquid regime may break down in the carbon 
nanotubes, due to a variety of low-temperature 
instabilities\cite{bal,eg,kane,yo,tse}. When the 
effective phonon-mediated interactions are taken into account,
it can be shown that the superconducting correlations may grow
large upon suitable screening of the Coulomb interaction in 
the nanotubes\cite{LET}. This has raised the hopes that 
such correlations could be amplified by considering nanotube
geometries with enhanced electron-phonon couplings. The case
of the small-diameter nanotubes, with radius down to 
$0.2 \; {\rm nm}$, is particularly interesting in that 
respect, since the large curvature of the tubule increases 
significantly the coupling to the elastic modes of the nanotube 
lattice\cite{bene,dme}.

The transport properties of small-diameter nanotubes have been 
studied in the experiment reported in Ref. \onlinecite{tang}, 
where it has been claimed evidence for a superconducting 
transition at about $15 \; {\rm K}$. It is uncertain however 
whether the observed features should be ascribed to the presence 
of nanotubes with the (5,0) or the (3,3) geometry, both being
consistent with the measured nanotube radius. Moreover, it 
remains also unclear the physical meaning of the estimated 
transition temperature, as the nanotubes in the experiment are 
placed in the channels of a zeolite matrix that does not allow 
coherent electron hopping between the nanotubes.

\begin{figure}
\includegraphics[height=6cm ]{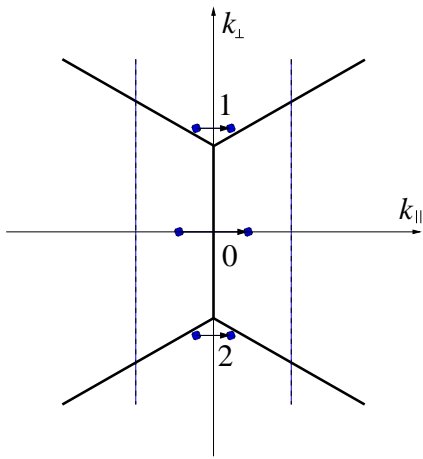}
\caption{Representation of the Brillouin zone of the (5,0) nanotubes 
showing the position of the Fermi points belonging to the low-energy
degenerate subbands (denoted by 1,2) and to the zero angular-momentum 
subband (denoted by 0).}
\label{brillo}
\end{figure}

\begin{figure}
\includegraphics[height=5cm ]{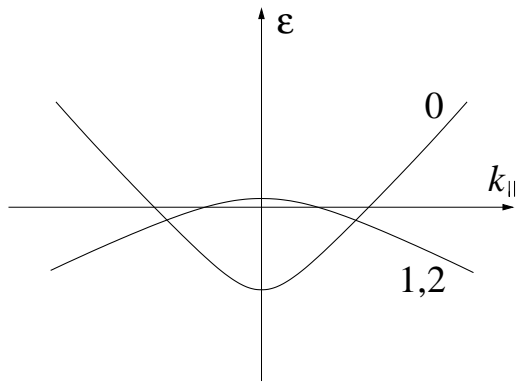}
\caption{Schematic representation of the dispersion of the low-energy 
degenerate subbands (superposed in the curve labelled by 1 and 2)
and the zero angular-momentum subband (labelled by 0) of the 
(5,0) nanotubes. The Fermi level corresponds to $\varepsilon = 0$
(in the undoped system).}
\label{threebnd}
\end{figure}

The aim of the present paper is to investigate the instabilites
that may lead to the breakdown of the Luttinger liquid 
in the small-diameter zig-zag nanotubes. We are going to focus
on the analysis of the (5,0) nanotubes, which deserve special
attention as they have three subbands crossing the Fermi level (in
the undoped system). The position of the different Fermi points
in momentum space is shown in Fig. \ref{brillo}. They correspond
to two degenerate subbands with opposite values of the angular
momentum around the nanotube axis (denoted by 1,2) and a third
subband with zero angular momentum (labelled by 0 in the figure).
The Fermi velocity is different in the nondegenerate and the 
degenerate subbands, being smaller in either case than for 
nanotubes of normal radius and leading to an enhanced density
of states at low energies. For the degenerate subbands, 
the Fermi velocity becomes
$v_F \approx 2.8 \times 10^5 \; {\rm m s}^{-1}$, while for the 
nondegenerate subband we have $v_F^{(0)} \approx 
6.9 \times 10^5 \; {\rm m s}^{-1}$. A diagram with the different 
shapes of the subbands along the momentum in the 
longitudinal direction is shown in Fig. \ref{threebnd}.

Together with the enhanced density of states at low energies,
the large curvature of the tubule makes the (5,0) geometry
the most appealing instance to study the effects of large
electronic correlations in the carbon nanotubes. We will apply 
renormalization group methods\cite{sol} to discern the low-energy 
instabilities of the electron system with three subbands at 
the Fermi level, extending the analysis carried out in Refs.
\onlinecite{eg} and \onlinecite{louie} for typical nanotubes 
with two low-energy subbands. This will allow us to study the 
growth of several response functions marking the tendency 
towards long-range order, including charge-density-wave (CDW) 
modulations with different values of the momentum and 
superconducting instabilities with different order parameters.

Our analysis will stress the relevance that the screening of 
the Coulomb interaction has for the development of different 
instabilities in the small-diameter nanotubes. For a single
nanotube, placed in general above some substrate, there is 
little reduction of the Coulomb potential, which remains 
long-ranged in the one-dimensional (1D)
system\cite{grab}. The Coulomb interaction 
has then a strong effective coupling, much larger than that of
the effective phonon-mediated interaction, so that the 
Coulomb repulsion plays the dominant role dictating 
the electronic properties about room temperatures. We will use 
bosonization methods\cite{emery} to achieve an exact treatment of 
the contributions from the Coulomb interaction at small
momentum-transfer, capturing the effects of the strong Coulomb 
repulsion at the nonperturbative level. The contributions from the 
effective phonon-exchange interactions will be analyzed next,
assessing their enhancement at low energies from the nontrivial 
scaling of the backscattering interaction channels.

On the other hand, we will pay also attention to the particular
conditions in the experiment reported in Ref. \onlinecite{tang}.
Each nanotube is embedded there in a large three-dimensional (3D)
array of nanotubes. Thus, the electrostatic coupling between 
charges in the different nanotubes leads to a large screening of 
the Coulomb potential. We will study this effect by using a kind
of generalized RPA scheme, taking into account the Coulomb 
interaction between all the nanotubes in the 3D array. We will 
show that, over distances much larger than the intertube 
separation, there is a regime where the nanotube array screens 
effectively as a 3D system, rendering short-ranged the Coulomb 
potential within each nanotube. This is the most important 
effect of the environment in the particular experimental setup 
described in Ref. \onlinecite{tang}. We will see that the 
strong reduction of the Coulomb interaction leads then to the 
prevalence of charge instabilities with similar character to 
the Wentzel-Bardeen singularity\cite{lm}.

We may compare the results of our investigation with those 
obtained for the small-diameter zigzag nanotubes by means of 
{\em ab initio} simulations\cite{ab} and mean-field-like 
calculations\cite{mf}. In Ref. \onlinecite{mf}, a superconducting 
instability has been found to be dominant in the (5,0) nanotubes, 
at a scale of about $1 \; {\rm K}$. On the other hand, a 
CDW instability has been identified in the same
geometry at room temperature in Ref. \onlinecite{ab}. Our results
stress that the scale and the type of electronic instability 
are very sensitive to the degree of screening 
applied to the nanotubes. We will find indeed that single 
(5,0) nanotubes (with just some screening from nearby 
gates) undergo the breakdown of the Luttinger liquid regime 
in favor of a CDW modulation at low energies. This is quite
different from the instability found in the 3D array of 
nanotubes. In either case, we will see that
the onset of the electronic 
instability takes place at temperatures ranging from 
$T_{c}\sim 10^{-4} \; {\rm K}$  up to $\sim 10^{2} \; {\rm K}$, 
depending on the dielectric constant of the environment.

\section{Luttinger liquid approach to electronic
properties}

As a first approximation to the electronic properties of the
small-diameter nanotubes, we begin our analysis by focusing
on the effects of the strong Coulomb repulsion. This is the
dominant interaction in scattering processes at low
momentum-transfer. Following Ref. \onlinecite{eg}, we take
the Coulomb potential in the wrapped geometry as 
\begin{equation}
V_C ({\bf r}-{\bf r}')=\frac{ e^{2} /\kappa }
 {\sqrt{(x-x')^{2}+4R^{2}\sin^{2}[(y-y')^{2}/2R^{2}]+a_{z}^{2} }}
\label{potent}
\end{equation}
where $a_{z} \simeq 1.6$ \AA and $R$ is the nanotube radius.
The effective strength of the interaction depends on the 
dielectric properties of the environment hosting the nanotubes. 
In general, we will encode the screening effects from external 
gates in the dielectric constant $\kappa $. We have to bare in 
mind, however, that this procedure is suitable for the description 
of single nanotubes, while it has to be appropriately improved 
for systems made of a manifold of nanotubes, as we will see in 
Sec. IV.

The Fourier transform $\tilde{V}_C (k, q) $ of the potential 
(\ref{potent}) reflects the long-range character of the interaction 
at small momentum-transfer. Thus, it has a logarithmic dependence 
at small longitudinal momentum $k$, which is characteristic of the 
1D Coulomb potential in momentum space\cite{sar}:
\begin{equation}
 \tilde{V}_C (k,0) \approx  
         \frac{2 e^2 }{\kappa }  \log(\frac{k_c+k}{k}) \, ,
\label{Coulomb}
\end{equation}
$k_c$ is in general of the order of the inverse of the 
nanotube radius $R$, as it is the memory that the electron system 
keeps of the finite transverse size, after projection of the 3D
potential onto the longitudinal direction of the tubule. The 
strength of the Coulomb interaction can be estimated from the 
dimensionless ratio $e^2 /\kappa v_F $, which for the 
small-diameter nanotubes is of the order of $\sim 8/\kappa$.
This gives a measure of the relevance of the Coulomb interaction,
which turns out to be in the strong-coupling regime in the
forward-scattering channels. 

Before going ahead, we proceed next to make a complete
catalogue of the different interaction channels. We first
borrow the classification that has been already made 
for metallic nanotubes with a pair of subbands at the Fermi 
level\cite{louie}. Thus, we introduce respective coupling constants
$g_i^{(j)}$ for the channels involving processes between
the two inner subbands of the small-diameter nanotubes. The lower
index discerns whether the interacting particles shift from one
subband to the other $(i=1)$, remain at different subbands $(i=2)$,
or they interact within the same subband $(i=4)$. The upper label
follows the same rule to classify the different combinations of
left-movers and right-movers. 

\begin{figure}
\includegraphics[height=9cm ]{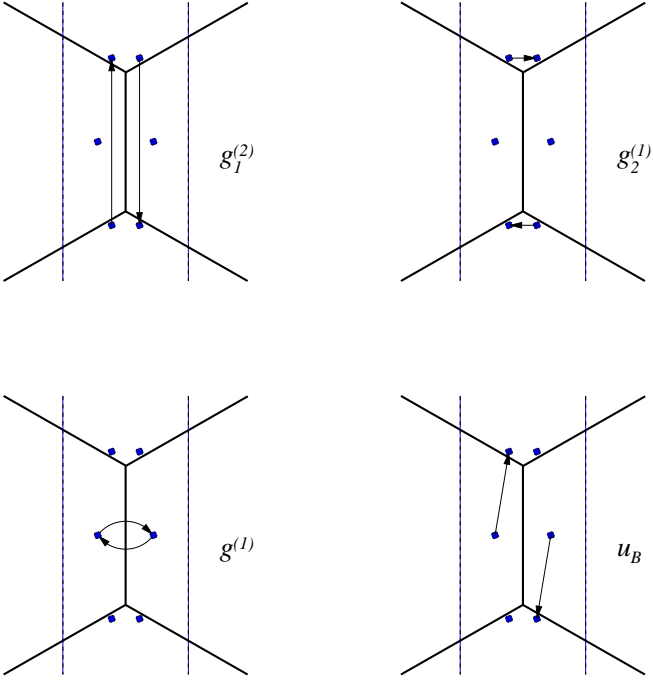}
\caption{Diagrams showing the shift in momentum space of the
interacting electrons and the momentum-transfer involved in different
backscattering processes in the (5,0) nanotubes.}
\label{exch}
\end{figure}

The above set of couplings $g_i^{(j)}$ has to be supplemented with
the couplings for the channels involving processes between 
a particle at one of the degenerate subbands and another at the 
nondegenerate subband. Now we can discern between interactions 
that keep each particle within its respective subband, to which we 
assign a coupling $f^{(2)} (f^{(4)})$ for particles of opposite (like) 
chirality, and backscattering interactions that lead to the exchange
of the subbands for the two particles, which we label with a coupling
$f^{(1)}$. Moreover, we have also the channels in which the particles
interact within the outer subband, and that we will
label as in the usual $g$-ology description, in terms of the
couplings $g^{(1)}, g^{(2)}$ and $g^{(4)}$. Finally, there are also
interaction processes in which two particles with about zero total
momentum exchange their position from the outer to the 
inner subbands, or vice versa, as depicted in Fig. \ref{exch}.
We will assign the coupling $u_F$ to that kind of 
interaction with no change of chirality of the particles, and
the coupling $u_B$ when there is a change in their chirality.
It is worthwhile to remark that, despite the role they
may have in the enhancement of some type of correlations, these 
interactions labelled by $u_F$ and $u_B$ have been overlooked in 
previous studies of the (5,0) nanotubes\cite{jap}.

Focusing again on the forward-scattering interactions, we 
may represent their contribution to the hamiltonian of the electron
system introducing the charge and spin density operators
\begin{eqnarray}
   \rho_{r i  }(x)  & = &
 \frac{1}{\sqrt{2}} 
    ( \Psi^{\dagger}_{r i \uparrow }(x) \Psi_{r i \uparrow }(x) +
 \Psi^{\dagger}_{r i \downarrow }(x) \Psi_{r i \downarrow }(x) ) \\
   \sigma_{r i  }(x)  & = &
 \frac{1}{\sqrt{2}}
 (   \Psi^{\dagger}_{r i \uparrow }(x) \Psi_{r i \uparrow }(x) -
 \Psi^{\dagger}_{r i \downarrow }(x) \Psi_{r i \downarrow }(x)  ) 
\end{eqnarray}
which correspond to the different electron fields
$\Psi_{r i \sigma } $ for the linear branches about the Fermi
points shown in Fig. \ref{threebnd}. 
We adopt a notation in which the index
$r = L, R$ is used to label the left- or right-moving character
of the linear branch, and the index $i = 0, 1, 2 $ to label the
subband. We will assume that the interaction mediated by 
phonon exchange (as well as the Coulomb interaction) does not 
depend on the spin of the interacting electrons, so that we will
carry out the discussion in terms of the charge density operators.

For the sake of organizing the different forward-scattering 
interactions, it is convenient to define the symmetric and 
antisymmetric combinations of corresponding density operators in 
the two degenerate subbands:
\begin{eqnarray}
\rho_{R \pm  }(k) & = &   \frac{1}{\sqrt{2}}
          (  \rho_{R 1  }(k) \pm  \rho_{R 2  }(k) )     \\
\rho_{L \pm  }(k) & = &   \frac{1}{\sqrt{2}}
          (  \rho_{L 2  }(k) \pm  \rho_{L 1  }(k)  )   \, .
\end{eqnarray}
With this change of variables, the hamiltonian for the
forward-scattering interactions can be written in the form
\begin{eqnarray}
H_{FS}  & = &    \frac{1}{2} v_F \int_{-k_c}^{k_c} dk
 \sum_{r=L,R} \sum_{i=\pm }   \rho_{r i } (k)
             \rho_{r i } (-k)     \nonumber    \\
  &  &    + \frac{1}{2} v_F^{(0)} \int_{-k_c}^{k_c} dk \;
   \sum_{r=L,R}   \rho_{r 0 } (k)    \;
             \rho_{r 0 } (-k)     \nonumber    \\
  &  &    + \frac{1}{2}  \int_{-k_c}^{k_c} \frac{dk}{2\pi } \;
    2   \left(   \rho_{R +} (k) \;
       (g_4^{(4)} + g_2^{(4)})     \;
   \rho_{R +} (-k)   \right.     \nonumber     \\
  &   &    +    \rho_{L +} (k) \;
       (g_4^{(4)} + g_2^{(4)})     \;
   \rho_{L +} (-k)           \nonumber     \\
  &  &    +   \rho_{R -} (k) \;
       (g_4^{(4)} - g_2^{(4)})     \;
   \rho_{R -} (-k)        \nonumber     \\
  &   &    +    \rho_{L -} (k) \;
       (g_4^{(4)} - g_2^{(4)})     \;
   \rho_{L -} (-k)           \nonumber     \\
  &  &    +  2  \rho_{R +} (k) \;
       (g_4^{(2)} + g_2^{(2)})     \;
   \rho_{L +} (-k)       \nonumber     \\
  &   &  \left.  +  2  \rho_{R -} (k) \;
       (g_2^{(2)} - g_4^{(2)})     \;
   \rho_{L -} (-k)   \right)   \nonumber    \\
  &   &  + \frac{1}{2}  \int_{-k_c}^{k_c} \frac{dk}{2\pi } \;
    2  \left(   \rho_{R 0 } (k) \;
       g^{(4)}     \;
   \rho_{R 0 } (-k)   \right.     \nonumber     \\
  &   &  \left.  +    \rho_{L 0 } (k) \;
       g^{(4)}    \rho_{L 0 } (-k)   \; 
         +  2  \rho_{R 0 } (k) \;
   g^{(2)}  \rho_{L 0 } (-k)  \right)   \nonumber  \\
  &   &  + \frac{1}{2}  \int_{-k_c}^{k_c} \frac{dk}{2\pi } \;
    4  \sqrt{2} \left(   \rho_{R + } (k) \;
       f^{(4)}     \;
    \rho_{R 0 } (-k)   \right.     \nonumber     \\
  &   &    +    \rho_{L + } (k) \;
       f^{(4)}    \rho_{L 0 } (-k) 
      +    \rho_{R + } (k) \; 
       f^{(2)}  \rho_{L 0 } (-k)  \nonumber  \\
  &  &  \left.   + \rho_{L + } (k)  \;
     f^{(2)}  \rho_{R 0 } (-k) \right) \, ,
\label{ham2}
\end{eqnarray}
where $k_c$ stands again for the momentum cutoff dictated by
the transverse size of the nanotube.

In order to diagonalize the hamiltonian (\ref{ham2}), we introduce
the fields $\Phi_+ (x), \Phi_- (x), \Phi_0 (x)$ (and their respective 
conjugate momenta, $\Pi_+ (x), \Pi_- (x), \Pi_0 (x)$) having
the following relation with the electron density operators:
\begin{eqnarray}
\partial_x \Phi_+ (x)  & = &  \sqrt{\pi }
           (  \rho_{L +}(x) + \rho_{R + }(x)  )        \\
\partial_x \Phi_- (x)  & = &  \sqrt{\pi }
           (  \rho_{L -}(x) + \rho_{R - }(x)  )        \\
\partial_x \Phi_0 (x)  & = &  \sqrt{\pi }
           (  \rho_{L 0}(x) + \rho_{R 0 }(x)  ) \, .
\end{eqnarray}
In terms of these fields, we can write the hamiltonian (\ref{ham2})
in the form
\begin{eqnarray}
H_{FS}   & = &   \frac{1}{2} u_+ \int dx
   \left(  K_+ ( \Pi_{+} (x)  )^2  +
  (1/K_+) ( \partial_x \Phi_{+}(x) )^2  \right)    \nonumber  \\
   &   &  +  \frac{1}{2} u_- \int dx  
    \left(  K_- ( \Pi_{-} (x)  )^2  +
    (1/K_-) ( \partial_x \Phi_{-}(x) )^2  \right)   \nonumber  \\
   &   &  +  \frac{1}{2} u_0 \int dx
   \left(  K_0 ( \Pi_{0} (x)  )^2  +
    (1/K_0) ( \partial_x \Phi_{0}(x) )^2  \right)   \nonumber  \\
   &   &   +  \frac{1}{2}  \int dx
    \frac{2  \sqrt{2}}{\pi}    \left(    \Pi_{+} (x) \;
       ( f^{(4)} - f^{(2)} )      \;
          \Pi_{0} (x)   \right.     \nonumber     \\
   &   &    \left.  +  \partial_x \Phi_{+} (x) \;
       ( f^{(4)} + f^{(2)} )      \;
         \partial_x \Phi_{0} (x)   \right) \, .
\label{hambf}
\end{eqnarray}
The renormalized velocities $u_+, u_-, u_0$ and charge
stiffnesses $K_+, K_-, K_0$ are given by the equations 
\begin{eqnarray}
 u_{\pm} K_{\pm}  & = &   
        v_F + (1/\pi ) \left( g_4^{(4)} \pm g_2^{(4)}
                - (g_2^{(2)} \pm g_4^{(2)}) \right)
                                             \label{v2}      \\
 u_{\pm} / K_{\pm}  & = &
     v_F +  (1/\pi )  \left( g_4^{(4)} \pm g_2^{(4)}
                + (g_2^{(2)} \pm g_4^{(2)}) \right)
                                              \label{v1}     \\
 u_{0} K_{0}  & = &
    v_F^{(0)} + (1/\pi ) \left( g^{(4)} - g^{(2)}  \right)
                                             \label{v02}      \\
 u_{0} / K_{0}  & = &
    v_F^{(0)} +  (1/\pi )  \left( g^{(4)} + g^{(2)}  \right) \, .
                                              \label{v01}    
\end{eqnarray}

At this stage, the hamiltonian (\ref{hambf}) can be brought to
diagonal form by: i) applying a simple canonical transformation 
\begin{eqnarray}
 \Phi_{+} =  \sqrt{K_+} \tilde{\Phi}_+
\,  &,&  \,
  \Pi_{+} = \frac{1} {\sqrt{K_+} }\tilde{\Pi}_+      \nonumber \\
 \Phi_{-} =  \sqrt{K_-} \tilde{\Phi}_-
\,  &,&  \,
  \Pi_{-} = \frac{1} {\sqrt{K_-} } \tilde{\Pi}_-      \nonumber  \\
 \Phi_{0} = \sqrt{K_0} \tilde{\Phi}_0
\,  &,&  \,
 \Pi_{0} = \frac{1} {\sqrt{K_0} }\tilde{\Pi}_0
\end{eqnarray}
and ii) making an additional rotation to disentangle the 
$(+,0)$ sector:
\begin{eqnarray} 
\tilde{\Phi}_{0} &=&  c_{0}\sqrt{\mu} \,  {\hat \Phi}_{0}
 +   s_{0} \sqrt{\nu} \, {\hat \Phi}_{+} \nonumber \\
 \tilde{\Pi}_{0} &=& \frac{c_{0}}{\sqrt{\mu}} \,{\hat \Pi}_{0} +
  \frac{s_{0}}{\sqrt{\nu}} \, {\hat \Pi}_{+} \nonumber      \\
 \tilde{\Phi}_{+} &=&  c_{+}\sqrt{\beta} \, {\hat \Phi}_{+}
 + s_{+} \sqrt{\alpha} \, {\hat \Phi}_{0} \nonumber \\
 \tilde{\Pi}_{+} &=& \frac{c_{+}}{\sqrt{\beta}} \,  {\hat \Pi}_{+} +  \frac{s_{+}}{\sqrt{\alpha}} \, {\hat \Pi}_{0} \, ,
\end{eqnarray}
where the parameters $c_0,s_0,c_+,s_+,\alpha, \beta, \mu, \nu$ are evaluated in
Appendix A. 
The important point is that, at the end, 
the excitations of the system governed by the hamiltonian 
(\ref{hambf}) are given by charge 
fluctuations, with velocities strongly renormalized 
by the interactions, and which constitute the whole spectrum
together with the spin fluctuations propagating with the 
unrenormalized velocities $v_F$ and $v_F^{(0)}$.

The picture that we have just developed corresponds to the 
Luttinger liquid regime of the electron system, in which
the main physical properties are given by the charge 
stiffnesses in the different sectors and by the velocities of
the charge and spin fluctuations\cite{emery}. 
These parameters may be used, 
for instance, to characterize the compressibilities in the 
symmetric, antisymmetric and zero angular-momentum sectors, 
which turn out to be proportional to the respective charge
stiffnesses.

The Luttinger liquid behavior in the carbon nanotubes  
holds at energies where there is approximate linear dispersion 
of the different subbands crossing the Fermi level.
This places typically an upper scale $E_c$ of the order of
$\sim 0.1 \; {\rm eV}$. In the case of single nanotubes, the 
dominant contribution to the forward-scattering couplings in 
(\ref{ham2}) comes from the strong Coulomb potential in 
(\ref{Coulomb}), with some infrared cutoff which is dictated
in general by the longitudinal size of the nanotube\cite{eg}. 
This leads to values of the charge stiffnesses that are 
typically smaller than 1 and in agreement with the experimental 
observations of Luttinger liquid behavior in the carbon 
nanotubes\cite{yao,exp}. For energy scales much lower than $E_c$, 
however, the rest of interaction channels neglected in the above 
analysis may become as important as the forward-scattering 
channels. This is possible as the strength of the 1D interactions 
depends in general on the energy scale of the interaction 
processes\cite{sol}. Electronic instabilities are then expected 
when some of the backscattering interactions leave the 
weak-coupling regime as the temperature or other relevant energy 
scale is lowered, as we analyze next in the case of the 
small-diameter (5,0) nanotubes.

\section{Backscattering interactions and low-energy
instabilities}

The small-diameter nanotubes are systems where the backscattering
interactions may lead to a significant enhancement of the electronic 
instabilities. Due to the large curvature of the tubes,  
the electron-phonon couplings turn out to be much larger than 
for nanotubes of typical radius. This leads consequently to a larger 
electron-electron interaction mediated by the exchange of 
phonons. In the case of single nanotubes, this effective interaction 
is anyhow overcome by the strong Coulomb repulsion given by the 
potential (\ref{Coulomb}) in the forward-scattering channels. The 
opposite situation is found however in the backscattering channels, 
as we discuss in what follows.

In general, the exchange of phonons between electronic currents 
gives rise to a retarded interaction, which can be represented by 
the effective potential obtained by integrating out the phonons in 
the many-body theory\cite{lm}. If we denote the electron-phonon 
couplings by $g_{ii'} (k)$ (labelling by $i,i'$ the subbands of the 
incoming and the outgoing electron), we can express the potential for 
the effective electron-electron interaction in the form
\begin{equation}
V_{ij,i'j'} (k,\omega ) = - 2g_{ii'}(k) g_{jj'}(-k)
  \frac{\omega_k }{- \omega^2 + \omega_k^2 } \, ,
\label{pot}
\end{equation}
where $\omega_k$ stands for the energy of the exchanged phonon with 
momentum $k$. The typical energy of the phonons in the optical 
branches (or in the acoustic branches at large momentum-transfer) is 
of the order of $\sim 0.1 \; {\rm eV}$ \cite{mf}, and therefore 
comparable to the energy cutoff $E_c$ of the 1D electron system. 
This allows us to take the interaction arising from (\ref{pot}) as 
a source of attraction, in the energy range where the 1D model 
makes sense.

The electron-phonon couplings in the (5,0) 
nanotubes have been analyzed with great detail in Ref. 
\onlinecite{mf}. It has been shown there that the couplings
with greater strength correspond to intraband processes, for which
the momentum-transfer is in the longitudinal direction. Following 
for instance the results of that reference for the backscattering 
processes within subband 1 (or subband 2), we find that the  
strength of the effective electron-electron interaction with 
$2 k_F$ momentum-transfer is given by the coupling 
\begin{equation}
 \sum_{\nu } 2 \left(  g_{11}^{(\nu )}  (2k_F) \right)^2 /
            v_F \omega_{2k_F}^{(\nu )}  \approx 0.6
\label{2k}
\end{equation}
after summing over the different phonon modes with nonvanishing
electron-phonon coupling\cite{mf}.
Similarly, for backscattering processes within the nondegenerate
subband we find the effective coupling 
\begin{equation}
 \sum_{\nu } 2 \left( g_{00}^{(\nu )} (2k_F^{(0)}) \right)^2 / 
    v_F \omega_{2k_F^{(0)}}^{(\nu )}   \approx 1.1 \, .
\label{2k0}
\end{equation}
The first of these values represents the contribution 
of the attractive phonon-mediated interaction to the couplings 
$g_2^{(1)}$ and $g_4^{(1)}$, while the second value provides the 
contribution to the coupling $g^{(1)}$. Making similar estimates of
the electron-phonon couplings and frequencies at large transverse
momentum-transfer, we have found that the couplings for the rest 
of backscattering interactions $g_{1}^{(1)}, g_{1}^{(2)}, f^{(1)}, 
u_F$ and $u_B$ are given by $\approx 0.5 v_F$.

To each of these backscattering contributions, one has to subtract 
an opposite correction from the Coulomb interaction. This is given
by the Fourier transform $\tilde{V}_C (k,q)$ of the potential in 
(\ref{potent}) at the corresponding momentum-transfer, with an 
additional factor of reduction that comes from the particular 
structure of the Bloch functions in the small-diameter 
nanotubes\cite{mf}. In general, the larger contributions from the 
Coulomb potential are found in the channels with the smaller
momentum-transfer. The forward-scattering channels deserve special
consideration, as the potential (\ref{Coulomb}) has to be evaluated
with a cutoff at small $k$. As already mentioned, this is provided 
in general by the length $L$ of the experimental samples, so that a 
sensible estimate is given typically by a spatial average of the 
potential 
$ \tilde{V}_C (k \approx 0,0)  \approx (2 e^2 /\kappa )  \log(k_c /k_0) $,
with $k_0 \sim 1/L \sim 10^{-3} k_c$.

In this section we analyze the instabilities of single nanotubes,
so that we can disregard for the moment any kind of intertube
interaction. Then, taking into account the contributions from the 
phonon-exchange and the Coulomb interaction, we obtain the following 
values for the couplings in the different interaction channels:
\begin{eqnarray}
  g_{4}^{(4)}/v_{F}=g_{2}^{(4)}/v_{F}=
    g_{4}^{(2)}/v_{F}=g_{2}^{(2)}/v_{F} \approx 100/\kappa  
                                     & &            \label{f}       \\
   g^{(4)}/v_{F}=g^{(2)}/v_{F}=
          f^{(4)}/v_{F}=f^{(2)}/v_{F} \approx  100/\kappa   & &  \\
   g_{4}^{(1)}/v_{F}=g_{2}^{(1)}/v_{F}  \approx  -0.6+3.6/\kappa & & \\
    g^{(1)}/v_{F}  \approx   -1.1+1.1/\kappa               & &   \\
   f^{(1)}/v_{F}=u_{B}/v_{F}   \approx  -0.5+0.5/\kappa    & &     \\
      u_{F}/v_{F}     \approx     -0.5+0.3/\kappa          & &   \\
   g_{1}^{(1)}/v_{F}=g_{1}^{(2)}/v_{F} \approx -0.5+0.05/\kappa  &.&
\label{l}
\end{eqnarray}
We note again that the different contributions from the Coulomb 
interaction (with a $1/\kappa $ reduction in each case) correspond
to the pertinent choices of the momentum-transfer in the evaluation
of $\tilde{V}_C (k, q) $.

\begin{figure}
\includegraphics[height=7.5cm ]{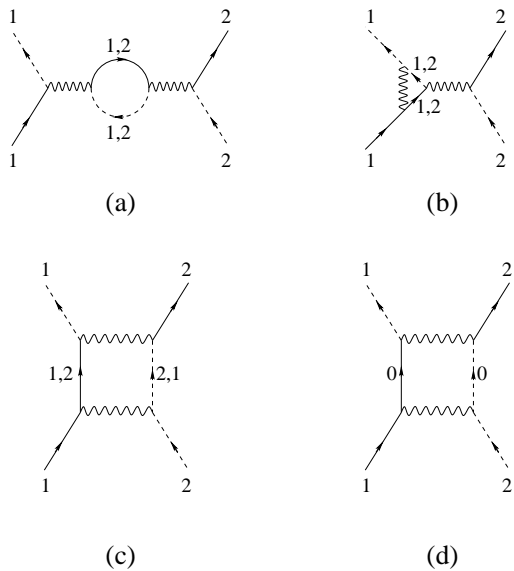}
\caption{Second-order diagrams with logarithmic dependence on the
frequency renormalizing the $g_2^{(1)}$ interaction. The full (dashed)
lines represent the propagation of electrons with right (left) 
chirality, and the wavy lines stand for the interactions. The labels
0,1,2 denote the respective low-energy subbands of the (5,0) 
nanotubes.}
\label{fig31}
\end{figure}

The relevance of the backscattering interactions comes from the 
fact that they give rise to quantum corrections that depend 
logarithmically on the energy scale. In the case of the $g_2^{(1)}$
interaction, for instance, the bare coupling is corrected to
second order by the diagrams shown in Fig. \ref{fig31}. 
By following the renormalization group program, we translate the 
logarithmic dependence on energy of the terms in the diagrammatic 
expansion into the scale-dependence of renormalized coupling 
constants\cite{sol}. Up to terms quadratic in the backscattering 
interactions, we find the complete set of scaling equations
\begin{eqnarray}
\frac{\partial g_1^{(1)}}{\partial l}  & = &
    - \frac{1}{\pi v_F}  (  g_1^{(1)} g_1^{(1)}
        +  g_1^{(2)} g_2^{(1)} )
        - \frac{1}{\pi v_F} \beta u_F u_B   \label{first}   \\
\frac{ \partial g_1^{(2)}}{\partial l} & = &
     (1 - \frac{1}{K_{-}}) g_1^{(2)}
     -   \frac{1}{\pi v_F}  ( g_2^{(1)} g_1^{(1)}   \nonumber  \\
     & &   +  (\beta /2)  ( u_F^2  + u_B^2 )  )         \\
\frac{ \partial g_2^{(1)}}{\partial l}  & = &
     (1 - \frac{1}{K_{-}})  g_2^{(1)}
   - \frac{1}{\pi v_F}  ( 2  g_4^{(1)} g_2^{(1)}  
                     -  g_4^{(1)} g_1^{(2)}    \nonumber     \\
   &  &   + g_1^{(2)} g_1^{(1)} + \beta u_F u_B )                 \\
\frac{ \partial g_2^{(2)}}{\partial l}  & = &
       -  \frac{1}{2 \pi v_F}  (  g_2^{(1)} g_2^{(1)}
           +  g_1^{(1)} g_1^{(1)}  +  g_1^{(2)} g_1^{(2)} )     \\
\frac{ \partial g_4^{(1)}}{\partial l}  & = &
       -  \frac{1}{\pi v_F}  (  g_4^{(1)} g_4^{(1)}
           +  g_2^{(1)} g_2^{(1)}  -  g_1^{(2)} g_2^{(1)} )     \\
\frac{ \partial g_4^{(2)}}{\partial l}  & = &
       -  \frac{1}{2 \pi v_F}  (  g_4^{(1)} g_4^{(1)}
           -  g_1^{(2)} g_1^{(2)}   )                     \\
\frac{\partial g^{(2)}}{\partial l}  & = &
    - \frac{1}{\pi v_F}  (  (\beta /2)    g^{(1)} g^{(1)} 
                   +    u_F^2  +  u_B^2    )               \\
\frac{\partial g^{(1)}}{\partial l}  & = &
    - \frac{1}{\pi v_F}  (  \beta   g^{(1)} g^{(1)}   
                      + 2  u_F  u_B    )                   \\
\frac{\partial f^{(2)}}{\partial l}  & = &
    - \frac{\alpha }{2 \pi v_F} ( f^{(1)} f^{(1)} - u_F^2 )  \\
\frac{\partial f^{(1)}}{\partial l}  & = &
    - \frac{\alpha }{\pi v_F} (  f^{(1)} f^{(1)} 
            +   u_B^2  -   u_F u_B  )                    \\
\frac{ \partial u_F}{\partial l}  & = &    \Delta u_F 
  -  \frac{1}{2 \pi v_F}  ( g_1^{(2)} u_F    \nonumber \\    
   &   &    
   + (g_2^{(1)}  +  g_1^{(1)} +  \beta g^{(1)}) u_B   )  \\
\frac{\partial u_B}{\partial l}  & = &    \Delta  u_B 
     -  \frac{1}{2 \pi v_F}  ( g_1^{(2)} u_B    \nonumber   \\
  &   &     + (g_2^{(1)}  +  g_1^{(1)}  + \beta g^{(1)}) u_F   )
                                                \nonumber    \\
  &  &  +  \frac{\alpha }{\pi v_F} f^{(1)} (u_F - 2u_B) \, ,
\label{last}
\end{eqnarray}
where $\beta = v_F / v_F^{(0)}$, $\alpha = 2/(1 + v_F^{(0)}/v_F)$
and the anomalous dimension $\Delta$ depends on $K_{+}, K_{-}, K_0$
and $f^{(2)}$ as shown in Appendix A. In the above equations, 
the variable $l$ stands for minus the logarithm of the energy
(temperature) scale measured in units of the high-energy scale
$E_c $ of the 1D model (of the order of $\sim 0.1 \; {\rm eV}$).

The equations for the $g_i^{(j)}$ couplings correspond to those
already written for a model with two subbands in Ref.
\onlinecite{louie}. Here we have incorporated a nonperturbative 
improvement of the equations by writing the exact dependence of 
the anomalous dimensions on the forward-scattering couplings 
through the $K_+, K_-$ and $K_0$ parameters. This is essential
to deal with the strong-coupling regime of the forward-scattering 
interactions in the case of single small-diameter nanotubes. 
Another important novelty is that the equations for the couplings 
in the degenerate and the nondegenerate subband mix through the 
couplings $u_F$ and $u_B$, which had been overlooked however in 
previous analyses of the (5,0) nanotubes.

In order to determine the electronic instabilities that may 
appear in single small-diameter nanotubes, we have 
solved the set of scaling equations (\ref{first})-(\ref{last}),
taking initial values for the couplings according to the above
discussion. The essential point is that the scaling equations 
reflect the strong screening that the couplings undergo at low 
energies, which leads them to a regime of large attraction. 
To find the dominant instability in the electron system, 
one has to look at the behavior of the different response 
functions $\chi$, which are catalogued in Appendix B. The regime 
of attraction leads to the large growth of some of the $\chi$; 
this points at a tendency towards long-range order, that cannot 
be completed anyhow at any finite frequency in the 1D system. 
The scaling of the interactions is cut off at some 
low-energy scale $\omega_c$ for which one of the charge 
stiffnesses $\beta, \mu$ or $K_-$  diverges. We characterize
the dominant instability by identifying the response function
that reaches the largest value at the scale $\omega_c$.

\begin{figure}
\includegraphics[height=5.5cm ]{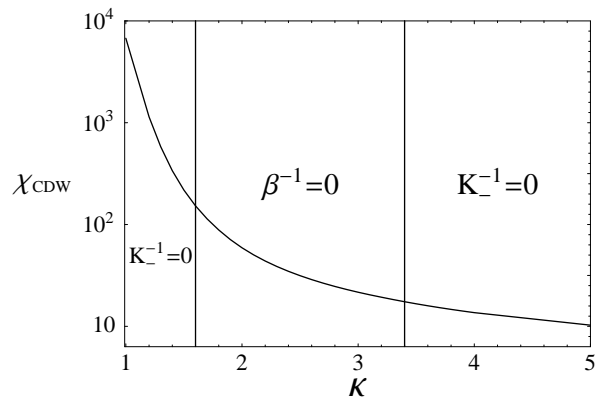}
\caption{Plot of $\chi_{CDW}(l_{c})$ as a function of $\kappa$.
Here $l_{c} = \log(E_c / \omega_c)$ is the scale at which the flow breaks down. 
The three regions with the corresponding divergent Luttinger liquid parameter 
are also indicated.}
\label{fig33a}
\end{figure}

By varying the dielectric constant, we have found two different regimes, 
namely a regime characterized by the divergence of $K_{-}$  (for $1< \kappa< 1.6$
and $\kappa > 3.4$), and a regime with  the divergence $\beta$ (for $1.6< \kappa< 3.4$). 
The largest growth for the respone functions 
corresponds to $\chi_{CDW}$, with large momentum 
connecting opposite Fermi points of the degenerate subbands
(see Appendix B).
We can understand this finding by looking at the 
competition between the phonon-exchange and Coulomb 
contributions in the different channels. At small $\kappa$,  
the couplings for large transverse momentum-transfer 
($g_{1}^{(1)}, g_{1}^{(2)}$) are 
attractive from the very beginning (see Eq.(\ref{l})). In particular, the flow is 
dominated by $g_{1}^{(1)}$, which is driven to large attraction, 
leading to a rapid growth of $\chi_{CDW}$.

Anyway, as shown in Fig.\ref{fig33a}, the $\chi_{CDW}$ grow
large only for small values of $\kappa$ ($\kappa \sim  1.4 \div 2$
for nanotubes in typical conditions).
At large $\kappa$ the maximum value for the all the respone functions 
remain close to 10, signaling that the tendency to ordering is weak. 

\begin{figure}
\includegraphics[height=5.5cm ]{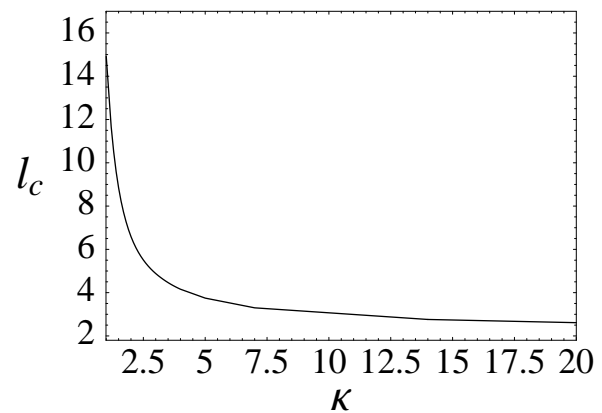}
\caption{Plot of minus the logarithm of the energy scale 
$l_{c} = \log(E_c / \omega_c)$ at which the flow breaks down, 
as a function of $\kappa$. }
\label{fig33}
\end{figure}

\begin{figure}
\includegraphics[height=7.0cm ]{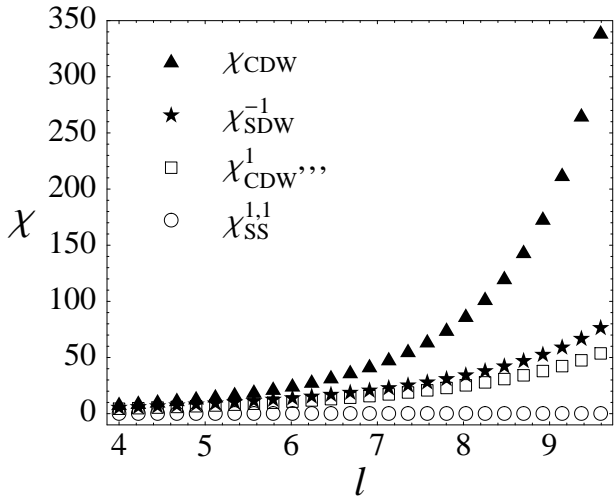}
\caption{Flow of the largest response functions 
(plus $\chi^{1,1}_{SS}$) at $\kappa=1.4$. }
\label{fig32}
\end{figure}

In Fig.\ref{fig33} we have plotted the critical value 
$l_{c} = \log(E_c / \omega_c)$ at which the flow breaks down 
as a function of $\kappa$. It appears that the energy scale 
$\omega_c$ becomes quite sensitive to the value of the dielectric 
constant, implying that the onset of the electronic 
instability can be found at temperatures ranging from 
$T_{c}\sim 10^{-4} \; {\rm K}$ (at $\kappa \approx 1$) up to a value 
$T_{c} \sim 10^{2} \; {\rm K}$ (reached from $\kappa \approx  10$). 
Finally, we observe from the comparison in Fig. \ref{fig32} that, 
while some charge and spin-density-wave response functions grow  
large by approaching the critical value $l_{c}$ (at least for small $\kappa$),
the superconducting 
response functions for different order parameters remain all small. 
This finding seems to rule out the possibility of having 
superconducting correlations in the (5,0) nanotubes, at least under 
the physical conditions considered in the present section. 

\section{Intertube screening and instabilities in a 3D
array of nanotubes}
\label{intertube}

While the above framework provides an accurate description of the 
instabilities of single small-diameter nanotubes, the analysis
has to be conveniently modified when the electron system is 
composed of a manifold of nanotubes. This is the 
situation in the experiments reported in Ref.
\onlinecite{tang}, where the samples contain large 3D arrays of
small-diameter nanotubes. Systems that are formed by the assembly
of a large amount of nanotubes, as is also the case of nanotube ropes,
require in general an additional analysis of the interactions arising
between different nanotubes. We will now focus on the conditions of
the experimental samples studied in Ref. \onlinecite{tang}, which
present a negligible intertube tunneling but have a
significant coupling between electronic currents in different
nanotubes.

Specifically, we will consider the case of an array of small-diameter 
nanotubes, arranged as a triangular lattice as viewed from a  
cross-section of the 3D array. We will denote by $d$ the distance
between nearest-neighbor nanotubes in such a lattice, having in mind
that, for the samples studied in Ref. \onlinecite{tang}, its value is 
given by $d \approx 1 \; {\rm nm}$. The coupling between different
nanotubes at positions ${\bf l}$ and ${\bf l'}$ (measured in the 
cross-section of the array) comes from the Coulomb interaction, due to 
its long-range character. It may couple currents with large 
transverse separation $| {\bf l}-{\bf l'}|$ as long as the longitudinal
momentum-transfer $k$ becomes as small as $| {\bf l}-{\bf l'}|^{-1}$. 
The Coulomb potential $V_{{\bf l}, {\bf l'}} (k)$ between currents in 
different nanotubes ${\bf l}$ and ${\bf l'}$ is actually obtained by
partially Fourier-transforming the 3D Coulomb potential
in the longitudinal direction: 
\begin{equation}
V_{{\bf l}, {\bf l'}} (k) \approx 
\frac{2 e^2}{\kappa } K_0 (| {\bf l}-{\bf l'}| k) \, ,
\label{intert}
\end{equation}
$K_{0}(x)$ being the modified Bessel function, 
which diverges logarithmically as $x \to 0$
and is exponentially suppressed as for $x>1$.

It is clear then that the nanotubes in the array may screen
efficiently the forward-scattering interactions within each
nanotube. This effect can be studied by extending the 
Random Phase Approximation (RPA) 
scheme to incorporate the electrostatic coupling between all the 
nanotubes in the array, as described in Appendix C. The main 
conclusion of this study is that the screened Coulomb potential 
$V^{(r)}_{{\bf l}, {\bf l}} (k) $ within each nanotube becomes finite in the limit 
of vanishing momentum $k \rightarrow 0$, reaching a saturation 
value $V^{(r)}_{{\bf l}, {\bf l}} (k = 0) \approx 0.08 e^2$. 
This corresponds to a dimensionless coupling 
$V^{(r)}_{{\bf l}, {\bf l}} (k = 0) /v_F 
\approx 0.65 $, which is about two orders of magnitude smaller than 
the one for single nanotubes. In this case the 
phonon-exchange contribution competes with the Coulomb repulsion 
also in the intratube forward-scattering channels, changing 
qualitatively the physical picture.

\begin{figure}
\includegraphics[height=2.5cm ]{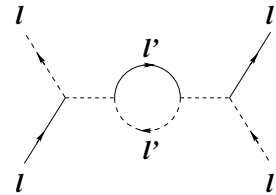}
\caption{Second-order process with logarithmic dependence on the
frequency renormalizing intraband interactions at a given nanotube
${\bf l}$ through the coupling with the nearest-neighbors ${\bf l'}$
in a 3D array of nanotubes. The dashed lines (without arrow) stand
for interactions between electronic currents in nearest-neighbor 
nanotubes.}
\label{fig41}
\end{figure}

\begin{figure}
\includegraphics[height=7cm ]{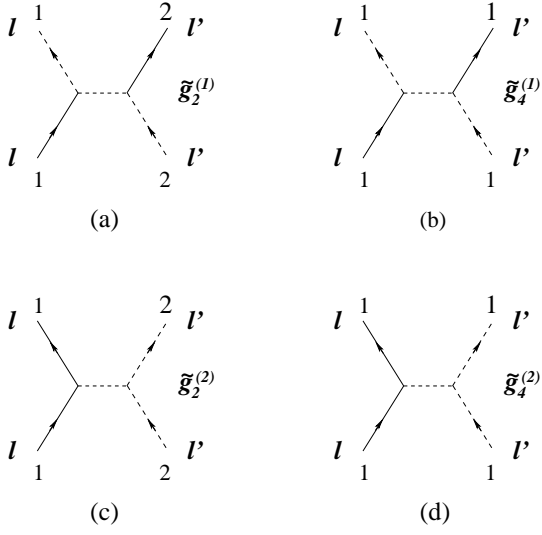}
\caption{Intertube interactions arising from the coupling between
electronic currents in nearest-neighbor nanotubes ${\bf l}$ and 
${\bf l'}$ of a 3D array. The meaning of the different lines is the
same as in Figs. \ref{fig31} and \ref{fig41}.}
\label{fig42}
\end{figure}

One more important effect is that the coupling between currents in 
neighboring nanotubes may also correct the intratube backscattering 
interactions with small momentum-transfer. This is due to the
fact that, for a longitudinal momentum-transfer $2k_F$, the Coulomb 
potential between nearest-neighbor nanotubes (such that $|{\bf
l}-{\bf l'}|=d$) still has a nonnegligible strength,
$V_{{\bf l}, {\bf l'}} (2k_F) \approx 0.037 e^2$. 
The important point is that this interaction gives rise to 
processes of the type depicted in Fig. \ref{fig41}, which depend on 
the energy scale and therefore introduce important renormalization 
effects for the intratube interactions at low energies. As observed 
from Fig. \ref{fig41}, the scaling analysis 
requires the definition of new backscattering interactions that 
couple electron currents in different nanotubes. We will assign
the couplings $\tilde{g}_2^{(1)}$ and $\tilde{g}_4^{(1)}$ to the
new channels represented in Figs. \ref{fig42}(a)-(b). 
As we are going to see
next, these new couplings mix upon renormalization with two more
channels corresponding to forward-scattering interactions between 
currents with different chirality in nearest-neighbor nanotubes.
These will be labelled by the couplings $\tilde{g}_4^{(2)}$ and 
$\tilde{g}_2^{(2)}$, as shown in Figs. \ref{fig42}(c)-(d).

\begin{figure}
\includegraphics[height=7.5cm ]{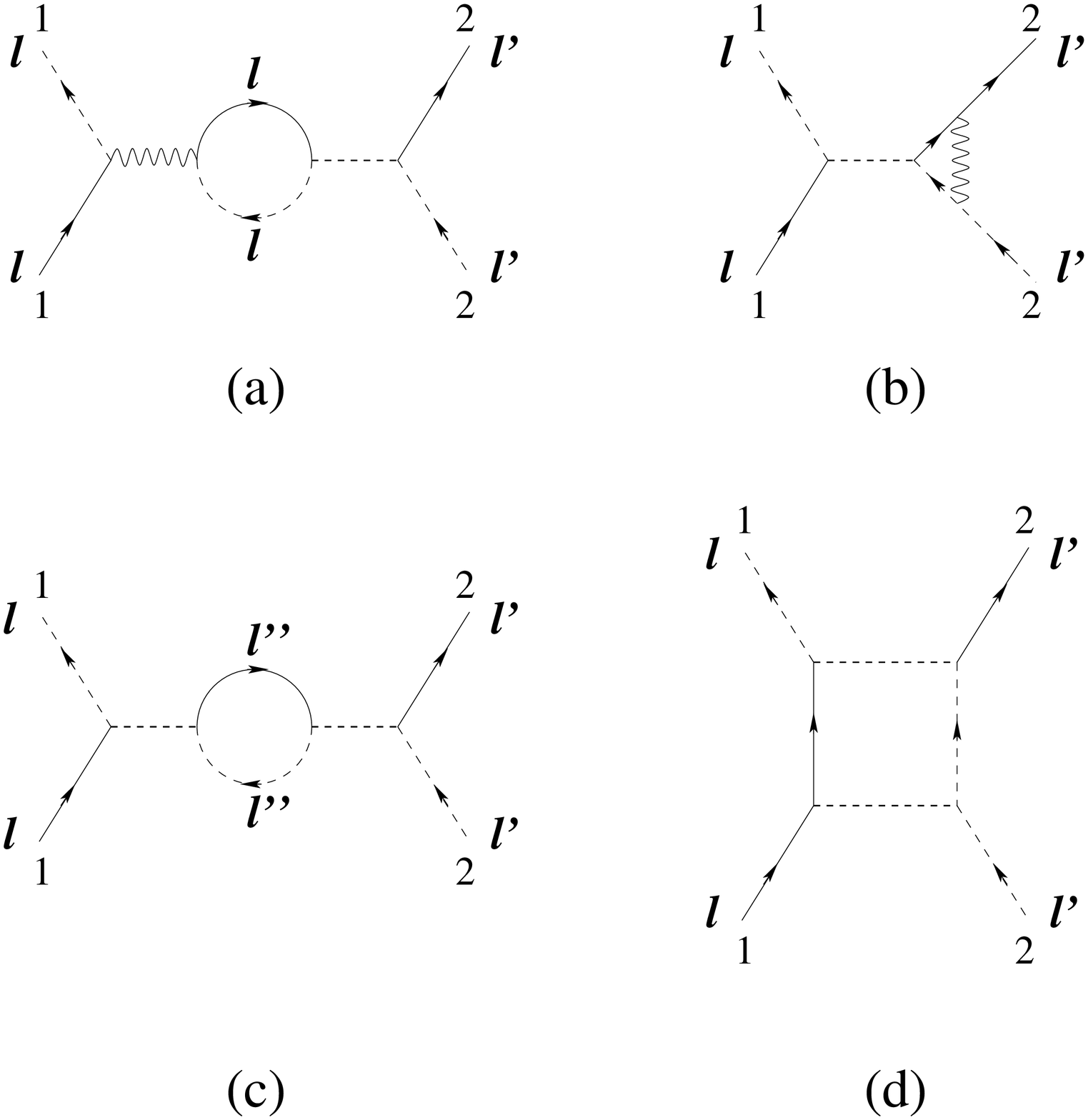}
\caption{Second-order diagrams with logarithmic dependence on the
frequency renormalizing the intertube $\tilde{g}_2^{(1)}$ interaction. 
The wavy lines stand for intratube interactions and the dashed lines
(without arrow) for interactions between nearest-neighbor nanotubes
${\bf l, l'}$ in a 3D array.}
\label{fig43}
\end{figure}

\begin{figure}
\includegraphics[height=8.5cm ]{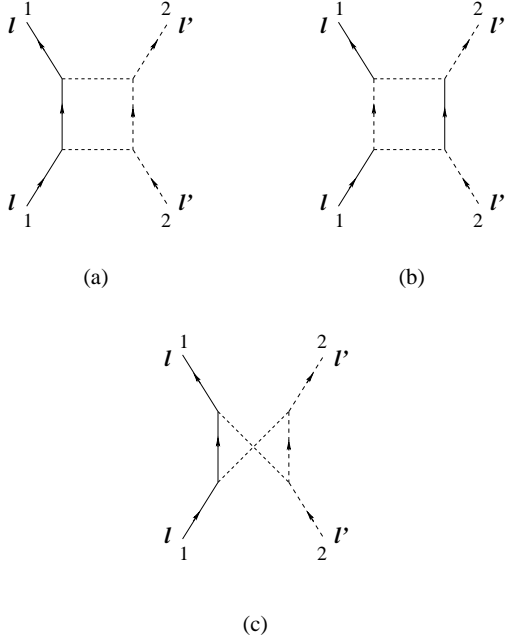}
\caption{Second-order diagrams with logarithmic dependence on the
frequency renormalizing the intertube $\tilde{g}_2^{(2)}$ interaction.
The dashed lines (without arrow) stand for interactions between 
nearest-neighbor nanotubes ${\bf l, l'}$ in a 3D array.}
\label{fig44}
\end{figure}

In order to determine the perturbative scaling of the new intertube
interactions $\tilde{g}_2^{(1)}$ and $\tilde{g}_4^{(1)}$, we pay 
attention to the second-order corrections that have 
logarithmic dependence on the energy scale. These have to be made of 
particle-particle processes or particle-hole loops with the nesting 
momenta $2k_F$ or $2k_F^{(0)}$. The different diagrams are shown in 
Fig. \ref{fig43}. We see that some of them involve the new intertube 
interactions $\tilde{g}_4^{(2)}$ and $\tilde{g}_2^{(2)}$. A consistent 
renormalization demands then the analysis of the scaling of this pair 
of couplings. This is dictated again by diagrams that depend 
logarithmically on the energy scale, depicted in 
Fig. \ref{fig44} to second order in perturbation theory.

The main role of the intertube backscattering interactions 
$\tilde{g}_4^{(1)}$ and $\tilde{g}_2^{(1)}$ is to screen the 
intratube backscattering interactions through processes of the type 
shown in Fig. \ref{fig41}. Thus, the scaling equations for $g_2^{(1)}$ 
and $g_4^{(1)}$ get new contributions in the 3D array of nanotubes, 
taking now the form
\begin{eqnarray}
\frac{ \partial g_2^{(1)}}{\partial l}  & = &
     (1 - \frac{1}{K_{-}})  g_2^{(1)}
   - \frac{1}{\pi v_F}  ( 2  g_4^{(1)} g_2^{(1)}
                     -  g_4^{(1)} g_1^{(2)}    \nonumber     \\
   &  &   + g_1^{(2)} g_1^{(1)} + \beta u_F u_B 
         + 12 \tilde{g}_4^{(1)}   \tilde{g}_2^{(1)}  )        \\
\frac{ \partial g_4^{(1)}}{\partial l}  & = &
       -  \frac{1}{\pi v_F}  (  g_4^{(1)} g_4^{(1)}
     +  g_2^{(1)} g_2^{(1)}  -  g_1^{(2)} g_2^{(1)}  \nonumber   \\
    &  &  + 6 \tilde{g}_4^{(1)} \tilde{g}_4^{(1)}  
          + 6 \tilde{g}_2^{(1)}     \tilde{g}_2^{(1)}  )  \, .  
\end{eqnarray}
We observe that the new terms are enhanced by a factor proportional
to the number of nearest-neighbors of each nanotube in the 3D array. 
The scaling equations for the new intertube couplings follow from the 
diagrams depicted in Figs. \ref{fig43} and \ref{fig44}:
\begin{eqnarray}
\frac{ \partial \tilde{g}_2^{(1)}}{\partial l}  & = &
 - \frac{1}{\pi v_F}  ( 2 g_4^{(1)} \tilde{g}_2^{(1)}
          +  2 g_2^{(1)} \tilde{g}_4^{(1)}     
       + 4  \tilde{g}_4^{(1)} \tilde{g}_2^{(1)}    \nonumber     \\
   &   &  +   \tilde{g}_2^{(2)}  \tilde{g}_2^{(1)}    
          -   g_4^{(2)}  \tilde{g}_2^{(1)}  
          -   g_1^{(2)}  \tilde{g}_4^{(1)}   )                  \\
\frac{ \partial \tilde{g}_4^{(1)}}{\partial l}  & = &
 - \frac{1}{\pi v_F}  ( 2 g_4^{(1)} \tilde{g}_4^{(1)}
          +  2 g_2^{(1)} \tilde{g}_2^{(1)}          \nonumber     \\
   &   &  \left.  + 2  \tilde{g}_4^{(1)} \tilde{g}_4^{(1)}   
         + 2  \tilde{g}_2^{(1)} \tilde{g}_2^{(1)}
          +   \tilde{g}_4^{(2)}  \tilde{g}_4^{(1)} \right. \nonumber    \\
   &   &   -   g_4^{(2)}  \tilde{g}_4^{(1)}  
           -   g_1^{(2)}  \tilde{g}_2^{(1)}    )                 \\
\frac{ \partial \tilde{g}_4^{(2)}}{\partial l}  & = &
 - \frac{1}{2 \pi v_F}  \tilde{g}_4^{(1)} \tilde{g}_4^{(1)} \label{g42}     \\
\frac{ \partial \tilde{g}_2^{(2)}}{\partial l}  & = &
 - \frac{1}{2 \pi v_F} \tilde{g}_2^{(1)} \tilde{g}_2^{(1)} \, .  \label{g22}
\end{eqnarray}

Let us now discuss the initial conditions for the new set of 
scaling equations. The intratube backscattering couplings have the 
same initial values as in the preceding section since, according to
the above discussion, the intertube screening in these channels 
is already incorporated in the scaling equations. Conversely,
in the forward-scattering channels, the screening may
be represented by finite diagrammatic corrections, which can be
summed up with the RPA approach described in Appendix C. As 
mentioned before, the Coulomb contribution is given by the 
dimensionless coupling $\tilde{V}^{(r)}_{{\bf l}, {\bf l}} (k = 0) /v_F 
\approx 0.65 $. The 
phonon-exchange contribution at vanishing momentum-transfer can
be estimated to be very similar to the value found from
(\ref{2k}), showing that there is now a substantial suppression
of the effects of the Coulomb repulsion in the intratube 
forward-scattering channels.

On the other hand, the intertube forward-scattering interactions 
are also screened due to the electrostatic coupling among 
the nanotubes in the array. These screening effects can be evaluated
again within the RPA scheme in Appendix C, with the result
that they render finite the intertube screened Coulomb potential
at $k \rightarrow 0$. Furthermore, the values of the intertube 
backscattering couplings $\tilde{g}_2^{(1)}$ and $\tilde{g}_4^{(1)}$ 
can be obtained from (\ref{intert}) with $|{\bf l}-{\bf l'} | = d$. The new initial 
conditions obtained in this way read:
\begin{eqnarray}
 \tilde{g}_{4}^{(2)}/v_F = \tilde{g}_{2}^{(2)}/v_F 
                                  \approx 0.002 \kappa   & &   \\ 
\tilde{g}_{2}^{(1)}/v_F = \tilde{g}_{4}^{(1)}/v_F  
                                     \approx 0.28/\kappa  &.&
\end{eqnarray}
We observe that the $\kappa$-dependence of the screened 
forward-scattering couplings is qualitatively different with respect 
to the nonscreened interactions. In particular, we find that the 
intertube couplings have an approximate linear dependence on 
$\kappa $, at least up to $\kappa \sim 5$ (see Appendix C).

\begin{figure}
\includegraphics[height=6.0cm ]{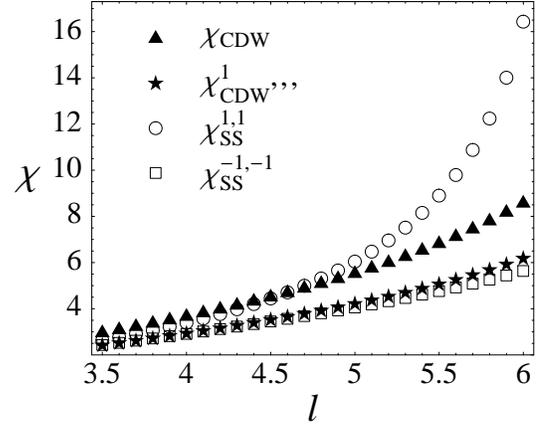}
\caption{Flow of the largest response functions  at $\kappa=2$.}
\label{fig45}
\end{figure}

The numerical integration of the scaling equations shows now a 
new physical regime, where the instability of the system is not 
characterized by the large growth of any of the response functions, 
but  only by the divergence of the charge stiffness $\beta$.  
We have plotted in Fig. \ref{fig45} a typical flow of the dominant
response functions up to the energy scale at which the divergence of
the charge stiffness occurs. We note that the screening effects produced
by the environment of nanotubes suppress the tendency to CDW ordering,
and enhance the superconducting correlations. Anyway, the values of
$\chi^{1,1}_{SS}$ do not grow large, making very unlikely that they may
give rise to any observable feature.

\begin{figure}
\includegraphics[height=5.5cm ]{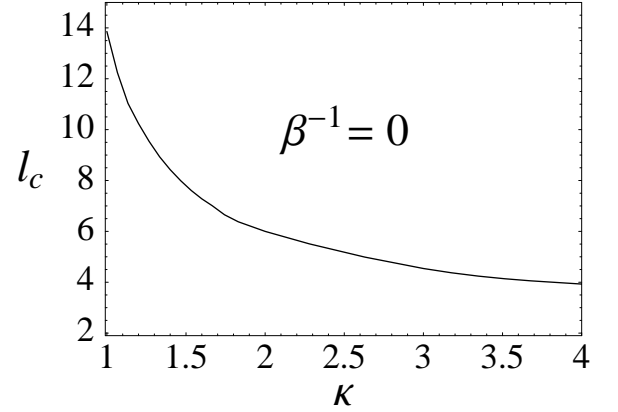}
\caption{Plot of minus the logarithm of the energy scale 
$l_{c} = \log(E_c / \omega_c)$ at which the flow breaks down, 
as a function of $\kappa$. The divergent Luttinger liquid 
parameter is also indicated.}
\label{fig46}
\end{figure}

Otherwise, the divergence of $\beta$ signals the onset of
a regime of strong attraction, as it implies the divergence of the 
compressibility and density-density correlator in the 
corresponding sector and the vanishing of the renormalized Fermi velocity
$ u_{+}/ \beta$ as well. This points at the development of 
phase separation into spatial regions with different electronic 
density, in close analogy with the physical interpretation of the 
Wentzel-Bardeen singularity. The critical energy scale for this 
instability is shown in Fig. \ref{fig46}. For the zeolite matrix of 
Ref. \onlinecite{tang}, the estimate of the dielectric constant 
gives $\kappa \approx 2 \div 4$, which corresponds to a transition 
temperature $T_{c} \approx 3 \; {\rm K} \div 20 \; {\rm K}$, 
in qualitative agreement with the value $T_{c} \approx 15 \; 
{\rm K}$ observed experimentally.

\section{Conclusion}

In this paper we have studied the electronic instabilities of the 
small-diameter (5,0) nanotubes, by analyzing the competition between
the Coulomb interaction and the effective interaction mediated by
the exchange of phonons. We have built our framework on the basis
that the Luttinger liquid is what characterizes the normal state of
the carbon nanotubes. This is a key assumption, since the low-energy
excitations of the 1D electron system are given in the normal state 
by charge and spin fluctuations, with absence of electron 
quasiparticles in the spectrum. This is consistent with the power-law
suppression of the conductance and the differential conductivity 
that has been measured in several transport experiments on carbon
nanotubes\cite{yao,exp,bach}. 
The deviations from this picture have to be understood 
as perturbations of the Luttinger liquid behavior, which in general
are nominally small as the strong Coulomb potential at small
momentum-transfer provides the dominant interaction in the system.

There are however instances in which the enhancement of the 
backscattering interactions (corresponding to some nesting momentum
between Fermi points) leads to the breakdown of the Luttinger
liquid picture. This certainly happens at sufficiently small 
temperature, since the backscattering interactions are amplified 
at low energies until they reach the strong-coupling 
regime\cite{sol}. There are also experimental conditions in which 
the strong Coulomb repulsion is largely screened. The screening 
effects are particularly important when the nanotubes form large 
assemblies\cite{sel}. This should explain for instance the 
appearance of a regime with superconducting correlations at low 
temperature in massive ropes. In this context, the increase in the 
curvature of the small-diameter nanotubes has to give rise to larger 
electron-phonon couplings\cite{bene}, 
implying in turn a natural enhancement of the backscattering 
interactions mediated by phonon-exchange. Thus, there have been 
good prospects to observe the effects of large electron correlations 
in the small-diameter nanotubes, specially in the (5,0) geometry 
which leads to a relatively large density of states at the Fermi 
level.

We have shown that the level of screening of the Coulomb 
interaction plays an important role in the development of the 
low-energy instabilities in the (5,0) nanotubes. In the case of 
single nanotubes exposed to the screening by external gates, 
the Luttinger liquid is unstable against the onset of a 
strong-coupling phase with very large CDW correlations. The 
temperature of crossover to the new phase depends crucially on
the dielectric constant $\kappa $ of the environment, ranging from 
$T_{c}\sim 10^{-4} \; {\rm K}$ (at $\kappa \approx 1$) up to a 
value $T_{c} \sim 10^{2} \; {\rm K}$ (reached from $\kappa \approx  
10$). The inspection of the response functions measuring the 
superconducting correlations shows that these do not have an 
appreciable growth in the regime where the CDW correlations 
grow large. Thus, even under conditions of strong screening by 
external charges, the observation of any superconducting 
feature seems to be excluded in single nanotubes with the 
(5,0) geometry.

The physical picture is different when we consider the case
of a large array of nanotubes, of the kind patterned by the
channels of the zeolite matrix in the experiments of Ref.
\onlinecite{tang}. The coupling between the nanotubes in the 
array leads to a large reduction of the Coulomb potential, which 
is rendered finite at vanishing momentum-transfer. The Coulomb
interaction happens to be now in the weak-coupling regime, making
therefore easier the destabilization of the Luttinger liquid.
We have found however that, before the superconducting 
correlations may grow large, a singularity is reached in one of
the charge stiffnesses of the Luttinger liquid. This implies the
divergence of the compressibility in one of the charge sectors,
and it has therefore the same physical interpretation as the 
Wentzel-Bardeen singularity\cite{lm}. 
As the density-density correlator
in each charge sector is proportional to the charge stiffness,
the divergence of the latter leads to a phase where the system
prefers to form electron aggregates, leaving some other regions
with a defect of electron density. The possibility of a phase 
of this type has been already anticipated in Ref. \onlinecite{dme}.
We have established that such a phase is realized under the 
particular experimental conditions reported in Ref. 
\onlinecite{tang}, prevailing over any superconducting 
instability at the transition temperatures found in our analysis.

The divergence of the charge stiffness opens the possibility of
having a genuine phase transition in the (5,0) nanotubes, since
the above breakdown of the Luttinger liquid takes place without the 
formation of long-range order in a continuous order parameter.
We have seen that, for a choice of the dielectric constant 
in accordance with the experimental conditions described in Ref.
\onlinecite{tang}, the transition temperature is of the same
order as the temperature of the crossover to the pseudogap regime 
observed in that reference. It is worthwhile to stress that
the divergence of the charge stiffness leads
to the development of a pseudogap in the single-particle spectrum.
The density of states near the Fermi level for electrons in subband $i$
scales in the Luttinger liquid regime as
\begin{equation}
n_{i}( \varepsilon ) \sim  \varepsilon^{\alpha_{i} }
\label{dos}
\end{equation}
with $\alpha_{i}$ as given in Appendix A. Quite remarkably, 
the $I$-$V$ characteristics
reported in Ref. \onlinecite{tang} show the evolution of the curves
for decreasing temperature into a power-law behavior with an 
increasingly large exponent. This is in agreement with the prediction 
obtained from (\ref{dos}) when approaching the singularity in the 
charge stiffness. It becomes then plausible that the crossover to
the pseudogap regime reported in Ref. \onlinecite{tang} may 
correspond to the breakdown of the Luttinger liquid found in 
the present framework.

There are however other experimental features of the small-diameter 
nanotubes described in Ref. \onlinecite{tang} that cannot
be explained in terms of the electronic instabilities we have found
in the present paper. The strong diamagnetic signal measured at
low temperatures has been interpreted in that reference as a 
consequence of entering a superconducting regime in the 
small-diameter nanotubes. It remains open the question of whether
the armchair (3,3) nanotubes can be responsible for such a regime,
or for some other type of electronic instability that may account
for the strong diamagnetic signal, as some preliminary results 
seem to indicate\cite{af}. 
Other questions that deserve attention refer to 
the properties of the small-diameter nanotubes when eventually
assembled into other aggregates different to those considered in the 
present paper. It may be interesting to see the influence of 
intertube tunneling between small-diameter nanotubes packed to
form massive ropes. It will be crucial then to discern whether 
the zigzag (5,0) and the armchair (3,3) nanotubes lead to 
different phases, and whether they may support the appearance
of larger superconducting correlations than those observed in 
massive ropes made of typical nanotubes.

\section*{Acknowledgements}
The financial support of the Ministerio
de Educaci\'on y Ciencia (Spain) through grant
BFM2003-05317 is gratefully acknowledged.
E. P. was also supported by INFN grant 10068.

\section*{Appendix A: Diagonalization of $H_{FS}$ and evaluation of the anomalous
dimensions}

In this Appendix we diagonalize $H_{FS}$ by means of a generalized Bogoliubov
transformation. Since in Eq. (\ref{hambf}) the $\{ \Phi_{-} , \Pi_{-} \}$
 sector is already decoupled, it
is sufficient to operate the transformation by mixing only the $\tilde{\Phi}_{+},\tilde{\Phi}_{0}$ fields
and their corresponding conjugate momenta $\tilde{\Pi}_{+},\tilde{\Pi}_{0}$. Therefore we introduce the
new operators ${\hat \Phi}_{+},{\hat \Phi}_{0}, {\hat \Pi}_{+},{\hat \Pi}_{0} $ defined by:
\begin{eqnarray} 
\tilde{\Phi}_{0} \equiv  c_{0}\sqrt{\mu}  {\hat \Phi}_{0}
 +   s_{0} \sqrt{\nu} {\hat \Phi}_{+}
\,  &,&  \,
 \tilde{\Pi}_{0} \equiv \frac{c_{0}}{\sqrt{\mu}} {\hat \Pi}_{0} +
  \frac{s_{0}}{\sqrt{\nu}} {\hat \Pi}_{+} 
                                                                 \nonumber      \\
 \tilde{\Phi}_{+} \equiv  c_{+}\sqrt{\beta}  {\hat \Phi}_{+}
 + s_{+} \sqrt{\alpha}  {\hat \Phi}_{0}
\,  &,&  \,
 \tilde{\Pi}_{+} \equiv \frac{c_{+}}{\sqrt{\beta}}   {\hat \Pi}_{+} +  \frac{s_{+}}{\sqrt{\alpha}} 
{\hat 
\Pi}_{0},
\nonumber 
\end{eqnarray} 
where $c_{0(+)}\equiv \mathrm{cos} \varphi_{0(+)}$ and
$s_{0(+)}\equiv \mathrm{sin} \varphi_{0(+)}$ in order to ensure the standard commutation
relations
 $[{\hat \Phi_{0(+)}(x) },{\hat \Pi}(x')_{0(+)}  ]= i\delta (x-x')$.

The angles $\varphi_{0},\varphi_{+}$ and the parameters $\mu,\nu, \alpha,\beta$ 
depend on $K_{0}, K_{+}, u_0, u_+, f^{(2)}, f^{(4)}$, and must be 
determined by imposing that
{\it (i)} $[{\hat \Phi(x)}_{0(+)},{\hat \Pi}(x')_{+(0)}   ]=0 $,
{\it (ii)} $H_{FS}$ written in terms of the new fields gets diagonal,
{\it (iii)} the coefficients of $(\partial_{x}  {\hat \Phi}_{0(+)})^{2} $
and  $( {\hat \Pi}_{0(+)})^{2} $ are the same. 
The corresponding algebraic system has an analytical solution in closed form
under the condition $f^{(2)}=f^{(4)}$ \cite{foot}:
\begin{eqnarray}
&&\varphi_{0}=\frac{1}{2} \mathrm{arctan}\left( \frac{4  \sqrt{2 K_{0} K_{+}} f^{(2)} /\pi }
{ u_{+}^{3/2} / u_{0}^{1/2}- u_{0}^{3/2} / u_{+}^{1/2}  } \right) \nonumber \\
&&\varphi_{+}=-\varphi_{0} \nonumber \\
&&\mu =  \left( c_{0}^{2}+s_{0}^{2} u_{+}^{2}/u_{0}^{2}- 4  \sqrt{2 K_{0} K_{+}} f^{(2)} 
c_{0}s_{0}  u_{+}^{1/2}/\pi u_{0}^{3/2}     \right)^{-1/2} \nonumber \\
&&\beta =  \left( c_{0}^{2}+s_{0}^{2} u_{0}^{2}/u_{+}^{2}+ 4  \sqrt{2 K_{0} K_{+}} f^{(2)} 
c_{0}s_{0}  u_{0}^{1/2}/\pi u_{+}^{3/2}     \right)^{-1/2} \nonumber \\
&&\alpha = \mu  u_{+}/u_{0} \, ,   \nonumber \\
&&\nu =   \beta u_{0}/u_{+} \, . \nonumber
\end{eqnarray} 
The renormalized Fermi velocities of the new free-boson hamiltonian read:
\begin{eqnarray}
u_{+} &\rightarrow&  u_{+}/\beta \, , \\ \nonumber
u_{0} &\rightarrow& u_{0}/\nu  \, . \nonumber
\end{eqnarray}

Now we use this result to calculate the nonperturbative contribution $\Delta$   
to the scaling equations of $u_{F}$ and $u_{B}$. 
For instance, let us focus on $u_{F}$, whose  operator has the form
\begin{eqnarray}
H_{u_{F}}= u_{F} \sum_{\sigma , \sigma '}( 
\Psi^{\dagger}_{R 0\sigma}   \Psi_{R 1\sigma}  
\Psi^{\dagger}_{L 0\sigma '} \Psi_{L 2\sigma '}+ \nonumber \\  
\Psi^{\dagger}_{R 0\sigma}   \Psi_{R 2\sigma}  
\Psi^{\dagger}_{L 0\sigma '} \Psi_{L 1\sigma '}+ \mathrm{h.c.} ).
\end{eqnarray}
The scaling dimension of $H_{u_{F}}$  is readily evaluated 
with the bosonization technique, that allows us to express the fermion 
fields in terms of boson operators. For example let us bosonize 
one of the contributions to $H_{u_{F}}$:
\begin{eqnarray}
\Psi^{\dagger}_{R 0}   \Psi_{R 1}  
\Psi^{\dagger}_{L 0} \Psi_{L 2} \propto \mathrm{exp} [-\frac{i}{\sqrt{2}}(-\phi_{R0}+
\phi_{R1}+ \phi_{L0}- \phi_{L2} )] \;. \nonumber
\end{eqnarray}   
Now we can  read the anomalous dimension of this operator produced by
the forward scattering interactions by expressing the boson fields $\phi$
in terms of the operators which diagonalize $H_{FS}$. 
The connection of the $\phi$'s with the operators 
entering in the hamiltonian in Eq. (\ref{hambf}) is given by:
\begin{eqnarray}
&&\phi_{R1}=\frac{1}{2\sqrt{2}}(\Phi_{+}+\Phi_{-}+\theta_{+}+\theta_{-}) \nonumber \\
&&\phi_{R2}=\frac{1}{2\sqrt{2}}(\Phi_{+}-\Phi_{-}+\theta_{+}-\theta_{-}) \nonumber \\
&&\phi_{L1}=\frac{1}{2\sqrt{2}}(\Phi_{+}-\Phi_{-}-\theta_{+}+\theta_{-}) \nonumber \\
&&\phi_{L2}=\frac{1}{2\sqrt{2}}(\Phi_{+}+\Phi_{-}-\theta_{+}-\theta_{-}) \nonumber \\
&&\phi_{R0}=\frac{1}{2}(\Phi_{0}+\theta_{0}) \nonumber \\
&&\phi_{L0}=\frac{1}{2}(\Phi_{0}-\theta_{0}) \nonumber  \, ,
\end{eqnarray}
where the $\theta$ fields are related to the $\Pi$'s by:
\begin{eqnarray}
\Pi_{\pm (0)}= -\partial_{x} \theta_{\pm (0)}  \, . \nonumber
\end{eqnarray}

This allows us to write
\begin{eqnarray}
&&\Psi^{\dagger}_{R 0}   \Psi_{R 1}  
\Psi^{\dagger}_{L 0} \Psi_{L 2} \propto \mathrm{exp} [\frac{i}{\sqrt{2}}(\theta_{0}-
\frac{1}{\sqrt{2}}\theta_{+}-\frac{1}{\sqrt{2}} \theta_{-} )] \nonumber \\
&& =  \mathrm{exp} \left[i\hat{\theta}_{0}  \left( 
 \frac{c_{0}}{\sqrt{2 K_{0} \mu }}+ \frac{s_{0}}{2\sqrt{K_{+} \alpha }}
 \right)  \right.  \nonumber \\ 
&& +i\hat{\theta}_{+}  \left. \left( 
 \frac{s_{0}}{\sqrt{2 K_{0} \nu}}- \frac{c_{0}}{2\sqrt{ K_{+} \beta}}
 \right)-i\hat{\Phi}_{-} \frac{1}{2\sqrt{K_{-}}}
 \right] \, . \nonumber
\end{eqnarray}
Finally the anomalous dimension $\Delta$ of the operator $H_{u_{F}}$
is obtained by looking at the coefficients of the free fields
in the exponent:
\begin{eqnarray}
&&\Delta= 1-   \left( 
 \frac{c_{0}}{\sqrt{2 K_{0} \mu }}+ \frac{s_{0}}{2\sqrt{ K_{+} \alpha}}
 \right)^{2} \nonumber \\
&& -
 \left( 
 \frac{s_{0}}{\sqrt{2 K_{0}\nu }}- \frac{c_{0}}{2\sqrt{ K_{+}\beta}}
 \right)^{2}    -     \left( \frac{1}{2\sqrt{K_{-}}}
 \right)^{2} \, . \nonumber
\end{eqnarray} 

By means of the same reasoning, we can compute also the anomalous dimensions of the 
operators  $O_{\mu}^{P,Q}$ defining the two-particle correlation functions,
introduced in Appendix B.
They encode the nonperturbative corrections
to the scaling Eqs. (\ref{sccorr})  for the response functions $\chi$ and read:
\begin{eqnarray}
&&\Delta_{DW}= 1- \frac{K_{-}}{2}  -\frac{K_{+}}{2} 
\left(  c_{0}^2 \beta + s_{0}^2 \alpha \right) \, , \nonumber \\
&&\Delta_{DW'}= 1- \frac{1}{2K_{-}}  -\frac{K_{+}}{2} 
\left(  c_{0}^2 \mu + s_{0}^2 \nu \right) \, , \nonumber \\
&&\Delta_{DW''}= 1- K_{0} \left(  c_{0}^2 \mu + s_{0}^2 \nu \right) \, , \nonumber \\
&&\Delta_{DW'''}= 1- \frac{K_{-}}{8}-\frac{1}{8K_{-}} 
- 2 \left( 
\frac{\sqrt{K_{+} \beta} c_{0} }{4} +
\frac{\sqrt{K_{0} \nu } s_{0} }{2\sqrt{2}}
\right)^{2} -  \nonumber \\
&& 2 \left( 
\frac{\sqrt{K_{0} \mu } c_{0} }{2\sqrt{2}}-
\frac{\sqrt{K_{+} \alpha} s_{0} }{4}
\right)^{2}  
- 2 \left( 
\frac{c_{0} }{4 \sqrt{K_{+} \beta}} -
\frac{ s_{0} }{2\sqrt{2}\sqrt{K_{0} \nu }}
\right)^{2}  \nonumber \\ 
&&-2 \left( 
\frac{s_{0} }{4 \sqrt{K_{+} \alpha}} +
\frac{ c_{0} }{2\sqrt{2}\sqrt{K_{0} \mu }}
\right)^{2}  \, , \nonumber \\
&&\Delta^{(a)}_{SC}= 1- \frac{1}{2K_{-}}  -\frac{1}{2K_{+}} 
\left(  \frac{c_{0}^2}{ \beta} + \frac{s_{0}^2}{ \alpha}  \right) \, , \nonumber \\
&&\Delta^{(b)}_{SC}= 1- \frac{1}{K_{0}} 
\left(  \frac{c_{0}^2}{ \mu} + \frac{s_{0}^2}{ \nu}  \right) \, , \nonumber \\
&&\Delta_{SC'}= 1-\frac{K_{-}}{2} - \frac{1}{2K_{+}} 
\left(  \frac{c_{0}^2}{ \mu} + \frac{s_{0}^2}{ \nu}  \right) \, , \nonumber \\
&&\Delta_{SC''}= 1- \frac{K_{-}}{8}-\frac{1}{8K_{-}} 
- 2 \left( 
\frac{\sqrt{K_{+} \beta} c_{0} }{4} -
\frac{\sqrt{K_{0} \nu } s_{0} }{2\sqrt{2}}
\right)^{2} -  \nonumber \\
&& 2 \left( 
\frac{\sqrt{K_{0} \mu } c_{0} }{2\sqrt{2}}+
\frac{\sqrt{K_{+} \alpha} s_{0} }{4}
\right)^{2}  
- 2 \left( 
\frac{c_{0} }{4 \sqrt{K_{+} \beta}} +
\frac{ s_{0} }{2\sqrt{2}\sqrt{K_{0} \nu }}
\right)^{2}  \nonumber \\ 
&&- 2 \left( 
\frac{s_{0} }{4 \sqrt{K_{+} \alpha}} -
\frac{ c_{0} }{2\sqrt{2}\sqrt{K_{0} \mu }}
\right)^{2}  \, . \nonumber 
\end{eqnarray} 

Finally we evaluate the anomalous dimensions $\alpha_{i}$
governing the density of states near the Fermi level
for electrons in the different subbands $i = 0, 1, 2$, 
in the Luttinger liquid regime:
\begin{eqnarray}
&& \alpha_{1}=\alpha_{2}= \frac{1}{8}  
\left[
K_{+}\left(  c_{0}^2 \beta + s_{0}^2  \alpha  \right)  \right. \nonumber \\ 
&& \quad \quad  \left. +
\frac{1}{K_{+}} \left(  \frac{c_{0}^2}{ \beta} + \frac{s_{0}^2}{ \alpha}  \right)
+ K_{-} + \frac{1}{K_{-}} -4
\right]  \, , \nonumber \\
&& \alpha_{0}= \frac{1}{4}  
\left[
K_{0}\left(  c_{0}^2 \mu + s_{0}^2  \nu  \right) +
\frac{1}{K_{0}} \left(  \frac{c_{0}^2}{ \mu} + \frac{s_{0}^2}{ \nu}  \right) -2
\right]  \, . \nonumber 
\end{eqnarray}

\section*{Appendix B: Scaling equations for the response functions }

\begin{figure}
\includegraphics[height=5.5cm ]{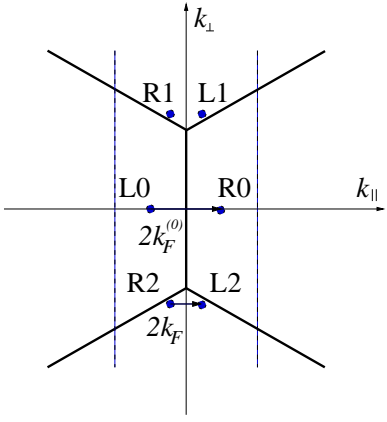}
\caption{Sketch of the Brillouin Zone of the (5,0) nanotube: 
the six Fermi points are indicated
with the labels of corresponding chirality and subband. The longitudinal 
momentum transfers $2k_{F}$ and $2k_{F}^{(0)}$ are also drawn.}
\label{brillo2}
\end{figure}

In order to study the competition between the possible
instabilities, one has to consider the scaling flow of the two-particle
correlation functions $\chi$. They measure the strength of the
quantum  fluctuations for a given type of ordering induced  by the pair-field
$O_{\mu}^{P,Q}$, in such a way that
\begin{eqnarray}
\chi_{\mu}^{P,Q}(k, \omega_m)=-\int_{0}^{\beta}\int_0 ^{L} d\tau d x e^{ikx -i\omega_m \tau} \nonumber \\
\times
\langle O_{\mu}^{P,Q}(x, \tau)^{\dagger} O_{\mu}^{P,Q} (0,0)   \rangle \, ,
\label{corfun}
\end{eqnarray}
where $L$ is the length of the nanotube, $\omega_m$ is
the bosonic Matsubara frequency and $P,Q = \pm 1$
take into account the band-entanglement\cite{caron}.
By neglecting the dependence on $\omega_m$ and expressing 
the momenta according to Fig.(\ref{brillo2}),  the complete
collection of the Fourier transforms of the response functions reads:
\begin{eqnarray}
&& O_{DW,\mu}^{P}(k \approx 2k_{F_{L2}})=\frac{1}{2\sqrt{L}} \sum_{p,\alpha,\beta} \left[
\Psi^{\dagger}_{R1 \alpha}(p-k) \sigma^{\alpha, \beta}_{\mu} \Psi_{L2 \beta} (p) \right. \nonumber \\
&& \quad \quad \quad  +P  \left. 
\Psi^{\dagger}_{L1 \alpha}(p-k) \sigma^{\alpha, \beta}_{\mu} \Psi_{R2 \beta}(p) \right] \, , \nonumber \\
&& O_{DW',\mu}^{P}(k \approx 2k_{F})=\frac{1}{2\sqrt{L}} \sum_{p,\alpha,\beta} \left[
\Psi^{\dagger}_{R1 \alpha}(p-k) \sigma^{\alpha, \beta}_{\mu} \Psi_{L1 \beta} (p) \right. \nonumber \\
&& \quad \quad \quad  +P  \left. 
\Psi^{\dagger}_{R2 \alpha}(p-k) \sigma^{\alpha, \beta}_{\mu} \Psi_{L2 \beta}(p) \right] \, , \nonumber\\
&& O_{DW'',\mu}(k \approx 2k_{F}^{(0)})=\frac{1}{\sqrt{L}} \sum_{p,\alpha,\beta} \left[
\Psi^{\dagger}_{R0 \alpha}(p-k) \sigma^{\alpha, \beta}_{\mu} \Psi_{L0 \beta} (p) \right] \, , \nonumber \\ 
&& O_{DW''',\mu}^{P}(k \approx k_{F_{L0}} -k_{F_{R1}})= \nonumber \\
&& \quad \quad \quad  \frac{1}{2\sqrt{L}} \sum_{p,\alpha,\beta} \left[
\Psi^{\dagger}_{R1 \alpha}(p-k) \sigma^{\alpha, \beta}_{\mu} \Psi_{L0 \beta} (p) \right. \nonumber \\
&& \quad \quad \quad +P  \left. 
\Psi^{\dagger}_{R0 \alpha}(p-k) \sigma^{\alpha, \beta}_{\mu} \Psi_{L2 \beta}(p) \right] \, , \nonumber \\ 
&& O_{SC,\mu}^{P,Q}(k \approx 0)=\frac{1}{\sqrt{3 L}} \sum_{p,\alpha,\beta} \left[
\Psi_{R1 \alpha}(-p+k) \sigma^{\alpha, \beta}_{\mu} \Psi_{L2 \beta} (p) \right. \nonumber \\
&& \quad \quad \quad +P  \left. 
\Psi_{R2 \alpha}(-p+k) \sigma^{\alpha, \beta}_{\mu} \Psi_{L1 \beta}(p)  \right. \nonumber \\
&& \quad \quad \quad  \left.  +Q \Psi_{R0 \alpha}(p-k) \sigma^{\alpha, \beta}_{\mu}
 \Psi_{L0 \beta}(p) \right] \, , \nonumber \\
&& O_{SC',\mu}(k \approx k_{F_{R1}}+k_{F_{L1}})=\nonumber \\
&&\quad \quad \quad   \frac{1}{\sqrt{L}} \sum_{p,\alpha,\beta} \left[
\Psi_{R1 \alpha}(-p+k) \sigma^{\alpha, \beta}_{\mu} \Psi_{L1 \beta} (p) \right]  \, , \nonumber \\
&& O_{SC'',\mu}(k \approx k_{F_{R1}}+k_{F_{L0}})=\nonumber \\
&& \quad \quad \quad  \frac{1}{\sqrt{L}} \sum_{p,\alpha,\beta} \left[
\Psi_{L0 \alpha}(-p+k) \sigma^{\alpha, \beta}_{\mu} \Psi_{R1 \beta} (p) \right] \;,  \nonumber 
\end{eqnarray}
where for density wave (DW) operators $\mu=0$ stands for CDW and $\mu=1,2,3$ for SDW; while for superconducting (SC)
operators $\mu=0$ stands for SS and $\mu=1,2,3$ for TS;  $ \sigma^{\alpha, \beta}_{\mu} $
are the Pauli matrices, with $ \sigma^{\alpha, \beta}_{0}= {\bf 1}_{2 \times 2} $.

The response functions $\chi$ defined in Eq. (\ref{corfun}) do not obey scaling
equations\cite{sol}, but the auxiliary functions defined as
$ {\bar \chi}(k,l)= \pi v_F (d/dl)\chi(k,l)$ do.
Finally the scaling equations for the ${\bar \chi}$ functions read:
\begin{eqnarray}
&&\frac{\partial}{\partial l}\mathrm{ln} {\bar \chi}_{CDW} = \Delta_{DW} 
- \frac{2}{\pi v_{F}} g_{1}^{(1)}  \, , \nonumber \\
&&\frac{\partial}{\partial l}\mathrm{ln} {\bar \chi}_{SDW} = \Delta_{DW} \, ,  \nonumber \\
&&\frac{\partial}{\partial l} \mathrm{ln} {\bar \chi}_{CDW'}^{P} = \Delta_{DW'} 
-\frac{2}{\pi v_{F}}
2 g_{4}^{(1)} + \frac{P}{\pi v_{F}}
( g_{1}^{(2)}-2 g_{2}^{(1)}) \, , \nonumber \\
&&\frac{\partial}{\partial l} \mathrm{ln} {\bar \chi}_{SDW'}^{P} = \Delta_{DW'} +
\frac{P}{\pi v_{F}}    g_{1}^{(2)} \, , \nonumber \\
&&\frac{\partial}{\partial l} \mathrm{ln} {\bar \chi}_{CDW''} = \Delta_{DW''} -\frac{2 \beta}{\pi v_{F}}
g^{(1)} \, , \nonumber \\
&&\frac{\partial}{\partial l}\mathrm{ln} {\bar \chi}_{SDW''} = \Delta_{DW''} \, , \nonumber \\
&&\frac{\partial}{\partial l}\mathrm{ln} {\bar \chi}_{CDW'''}^{P} = \Delta_{DW'''} -
\frac{2 \alpha}{\pi v_{F}} f^{(1)} +
 \frac{\alpha P}{\pi v_{F}} (u_{F} -2u_{B}) \, , \nonumber \\
&&\frac{\partial}{\partial l}\mathrm{ln} {\bar \chi}_{SDW'''}^{P} = \Delta_{DW'''}
 + \frac{\alpha P }{\pi v_{F}} u_{F}  \, ,  \nonumber \\
&&\frac{\partial}{\partial l}\mathrm{ln} {\bar \chi}_{SS}^{P,Q} = 
\frac{2}{3}\Delta^{(a)}_{SC}  +\frac{1}{3}\Delta^{(b)}_{SC} \nonumber \\
&& \quad \quad  + \frac{2}{3\pi v_{F}}
\left[- g_{1}^{(1)} -P(g_{1}^{(2)}+g_{2}^{(1)}) \right] \nonumber \\
&& \quad \quad   - \frac{\beta}{3\pi v_{F}}
\left[2(Q+PQ)( u_{F}+u_{B}) +g^{(1)} \right]  \, , \nonumber \\
&&\frac{\partial}{\partial l}\mathrm{ln} {\bar \chi}_{TS}^{P,Q} = 
\frac{2}{3}\Delta^{(a)}_{SC}  +\frac{1}{3}\Delta^{(b)}_{SC}  \nonumber \\
&& \quad \quad  + \frac{2}{3\pi v_{F}}
\left[ g_{1}^{(1)} -P(g_{1}^{(2)}-g_{2}^{(1)}) \right] \nonumber \\
&&\quad \quad   - \frac{\beta}{3\pi v_{F}}
\left[2(Q+PQ)( u_{F}-u_{B}) -g^{(1)} \right] \, , \nonumber \\
&&\frac{\partial}{\partial l}\mathrm{ln} {\bar \chi}_{SS'} = 
\Delta_{SC'} -
\frac{1}{\pi v_{F}} g_{4}^{(1)}  \, , \nonumber \\
&&\frac{\partial}{\partial l}\mathrm{ln} {\bar \chi}_{TS'} = 
\Delta_{SC'} +
\frac{1}{\pi v_{F}} g_{4}^{(1)} \, , \nonumber \\
&&\frac{\partial}{\partial l}\mathrm{ln} {\bar \chi}_{SS''} = 
\Delta_{SC''} -
\frac{\alpha}{\pi v_{F}}
 f^{(1)}  \, , \nonumber \\
&&\frac{\partial}{\partial l}\mathrm{ln} {\bar \chi}_{TS''} = 
\Delta_{SC''} +
\frac{\alpha}{\pi v_{F}} f^{(1)} \, , \label{sccorr}  
\end{eqnarray}
where the nonperturbative contributions encoded in the anomalous
dimensions $\Delta_{DW},\Delta_{DW'},\Delta_{DW''},\Delta_{DW'''},
\Delta^{(a,b)}_{SC},\Delta_{SC'},\Delta_{SC''}$ are evaluated
in Appendix A.

\section*{Appendix C: Intertube screening for forward scattering
processes }

In this Appendix we study the Coulomb interaction at small
momentum-transfer between
electrons belonging to different nanotubes in a 3D array. The
long-range intertube effects operate in two ways: {\it i)} it
provides a screening of the bare intratube forward-scattering couplings
$g_{2}^{(2)},g_{2}^{(4)},g_{4}^{(2)},g_{4}^{(4)},g^{(2)},g^{(4)}, 
f^{(2)}, f^{(4)}$, and {\it ii)} produces new intertube couplings 
$\tilde{g}_{2}^{(2)},\tilde{g}_{4}^{(2)}$ (see Eqs. (\ref{g42}), 
(\ref{g22})) with nontrivial scaling flow. Here we take into
account the screening effects by the nanotube environment by
generalizing the approach devised in Ref. \onlinecite{hawrylak}, which
essentially consists on a RPA treatment of the dielectric constant. The
RPA approximation is justified as long as the electrons are not allowed
to tunnel between different nanotubes, and we assume that this condition
applies to the system under consideration in the main body of the paper. 
It is important to stress
that the screened interactions obtained by this approach do not contain
any dependence on the energy cutoff. As a consequence, the RPA screening
only imposes the initial value of the forward-scattering couplings
in Section \ref{intertube}, but does not affect their scaling flow.

Let ${\bf l}$ and ${\bf l'}$ be the positions of two nanotubes measured
over a section of the 3D array, orthogonal to the tube axes. We
introduce the long-range bare Coulomb interaction with
momentum-transfer $k$ along the longitudinal direction 
by partially Fourier-transforming the 3D Coulomb potential:
\begin{equation} 
V_{{\bf l},{\bf l'}}(k)=\frac{2e^{2}}{\kappa} K_{0}(k|{\bf l}-{\bf
l'}|), 
\label{inter} 
\end{equation} 
where $K_{0}(x)$ is the modified
Bessel function, which diverges logarithmically as $x \to 0$
and is exponentially suppressed for $x>1$. It is worth to observe
that $V_{{\bf l},{\bf l'}}(k)$ does not take into account the
variation of the electronic Bloch wavefunctions around the waist
of the nanotubes. This approximation is justified as long as
the transverse momentum-transfer is small, i.e. for forward-scattering
processes. When $V_{{\bf l},{\bf l'}}(k)$ is evaluated within the
same nanotube, it is understood that $|{\bf l}-{\bf l'}|\sim R$,
and we recover the intratube interaction of
Eq. (\ref{Coulomb}), with the logarithimc divergence at small 
momentum-transfer.

The RPA-screened $V_{{\bf l},{\bf l'}}^{(r)}(k)$ obeys the Dyson equation
\begin{equation} 
V_{{\bf l},{\bf l'}}^{(r)}(k)=V_{{\bf l},{\bf
l'}}(k)+\Pi \,\sum_{{\bf l''}}V_{{\bf l},{\bf l''}}(k) \,
V_{{\bf l''},{\bf l'}}^{(r)}(k), 
\label{dyson} 
\end{equation}
where ${\bf l''}$ runs over all the positions of the nanotubes in the
3D array and $\Pi$ is the polarizability of the 1D electron
gas; its zero-frequency Fourier transform reads 
\begin{equation}
\Pi(k)=\frac{2}{L}\sum_{q}
\frac{f(\varepsilon_{q+k})-f(\varepsilon_{q})}
        {\varepsilon_{q+k}-\varepsilon_{q}},
\label{kin} 
\end{equation}
where $f$ is the Fermi-Dirac distribution and $L$ is the lenght of the
nanotubes. In order to solve Eq. (\ref{dyson}), we introduce the 2D
Fourier transforms of $V_{{\bf l},{\bf l'}}(k)$ and $V_{{\bf l},{\bf
l'}}^{(r)}(k)$:
\begin{equation} 
V_{{\bf l},{\bf l'}}(k)=\left( \frac{d}{2 \pi} \right)^{2} \int_{BZ}
d^{2}{\bf p} \, \phi(k,{\bf p}) \, e^{i {\bf p} \cdot ({\bf l}-{\bf l'})},
\label{fourier} 
\end{equation} 
and the same with $V \rightarrow V^{(r)}$ and $\phi \rightarrow \phi^{(r)}$,
where $BZ$ indicates the first Brillouin
zone of the nanotube lattice in the cross-section of the array, with lattice
constant $d\sim 1$ nm. In terms of $\phi(k,{\bf p})$ and $\phi^{(r)}(k,{\bf
p})$,  the Dyson equation reduces to 
\begin{equation} 
\phi^{(r)}(k,{\bf p})=\phi(k,{\bf p})+\Pi(k) \, \phi(k,{\bf p})\, \phi^{(r)}(k,{\bf p}).
\label{algdys} 
\end{equation} 
This permits to calculate the screened forward-scattering interaction 
with longitudinal momentum-transfer $k$ between electrons in tubes 
at ${\bf l}$ and ${\bf l'}$: 
\begin{equation}
V_{{\bf l},{\bf l'}}^{(r)}(k)=\left( \frac{d}{2 \pi}
\right)^{2}\int_{BZ} d^{2}{\bf p} \frac{\phi(k,{\bf p})}{1-\Pi (k)\phi(k,{\bf p})} 
 e^{i {\bf p}\cdot ({\bf l}-{\bf l'})}.
\label{screened} 
\end{equation} 

The numerical evaluation of $V_{{\bf l},{\bf l'}}^{(r)}(k)$
shows that such interaction is not negligible only for
${\bf l}={\bf l'}$ and for nearest-neighbor tubes, i.e. for $|{\bf
l}-{\bf l'}|=d$. This calls for the introduction of intertube couplings 
with $|{\bf l}-{\bf l'}|=d$. As discussed in Section \ref{intertube}, 
two of them ($\tilde{g}_{2}^{(2)}$ and $\tilde{g}_{4}^{(2)}$) have nontrivial 
scaling flow, and affect also the scaling of the intratube couplings.

The logarithmic divergence of $V_{{\bf l}, {\bf l'}} (k)$
at $k=0$ is cured by the RPA screening;
nevertheless the effective interactions are still enhanced at small $k$,
but they saturate at a finite value.
For the intratube interaction we have found that the saturation value is 
essentially independent on $\kappa$  and is 
$V_{{\bf l}, {\bf l}}^{(r)}(k = 0)/v_{F} \approx 0.65 $.  
In the intertube case, the RPA
screening is much more efficent and the saturation  value
for $|{\bf l}-{\bf l'}|=d$
is $V_{{\bf l}, {\bf l'}}^{(r)}(k = 0)/v_{F} \approx 0.002 \kappa$,
with an approximate linear dependence  on $\kappa$, at least up to 
$\kappa \sim 5$. 
As discussed above, these asymptotic values are used as initial conditions 
for the forward-scattering couplings in the scaling equations in 
Section \ref{intertube}.



\begin{thebibliography}{99}
\bibitem{foot}
This condition is satisfied with great approximation in the systems
considered in the paper, as the couplings $f^{(2)}$ and $f^{(4)}$
have equal values at the beginning of the scaling flow and they
are very softly renormalized at low temperatures.
\end{thebibliography}
\end{document}